# Resource-Driven Mission-Phasing Techniques for Constrained Agents in Stochastic Environments

**Jianhui Wu** JIANHUIW@UMICH.EDU
**Edmund H. Durfee** DURFEE@UMICH.EDU
*Computer Science and Engineering, University of Michigan*
*Ann Arbor, MI 48109 USA*

## Abstract

Because an agent's resources dictate what actions it can possibly take, it should plan which resources it holds over time carefully, considering its inherent limitations (such as power or payload restrictions), the competing needs of other agents for the same resources, and the stochastic nature of the environment. Such agents can, in general, achieve more of their objectives if they can use — and even create — opportunities to change which resources they hold at various times. Driven by resource constraints, the agents could break their overall missions into an optimal series of phases, optimally reconfiguring their resources at each phase, and optimally using their assigned resources in each phase, given their knowledge of the stochastic environment.

In this paper, we formally define and analyze this constrained, sequential optimization problem in both the single-agent and multi-agent contexts. We present a family of mixed integer linear programming (MILP) formulations of this problem that can optimally create phases (when phases are not predefined) accounting for costs and limitations in phase creation. Because our formulations simultaneously also find the optimal allocations of resources at each phase and the optimal policies for using the allocated resources at each phase, they exploit structure across these coupled problems. This allows them to find solutions significantly faster (orders of magnitude faster in larger problems) than alternative solution techniques, as we demonstrate empirically.

## 1. Introduction

An omnipresent issue in realistic application domains for autonomous agents is that agents are resource-limited. Resources enable action. For example, an agent with a camera can capture an image, an agent with a gripper can manipulate objects, an agent with an auxiliary battery pack can take more actions before it must recharge, and an agent with an additional memory chip can solve larger computational problems. Given the resources it possesses, an agent should utilize them to take the best sequences of actions that it can, depending on its objectives and environment.

In this paper, we consider the situation where agents have some degree of control over the resources they can choose to possess, subject to inherent (unavoidable) limitations of the agents themselves, as well as to contention over resources. Agents' inherent limitations stem from what we will call *capacity* constraints. For example, a mobile robot agent (say, a Mars Rover) might have weight limitations for the payload it can carry, such that it cannot carry a camera and gripper at the same time. Or the physical configuration of such resources might preclude some combinations, such as if the gripper arm necessarily obstructs the camera view. The power drawn across combinations of peripherals might exceed an agent's power supply, or the computational cycles demanded by a combination over a time interval might exceed the agent's processing power. In short, for any of a number of reasons, an agent might lack the capacity to effectively possess all of the resources that it might find useful, in which case it needs to determine which subset combination of resources will, in expectation, allow it to act most effectively over time, given uncertainty over the future evolution of its environment.

In a multiagent setting, an agent might fail to possess a potentially useful resource not only because of capacity constraints, but also because of resource scarcity constraints. For example, there might be fewer instruments of some type, such as working cameras or grippers, than there are robots (Mars Rovers), in which case a cooperative agent should only get one of these resources if it expects to make better use of it than other agents who would not get it. Along similar lines, if the number of licenses for running a particular piece of software are limited, then cooperative agents should allocate them in the best possible way for their collective benefit. Or, if the number of satellites to remotely control in order to acquire needed images is restricted, then assignments of these to agents should be done judiciously.

Dolgov and Durfee (2006) looked at these kinds of problems, studying efficient techniques by which agents can assess the value of alternative resource combination (bundle) assignments in terms of the execution policies (and the expected utilities of those policies) that the resources enable. That work focused on the question of finding an optimal static resource allocation for an agent (or multiple agents) where the agent(s) then make sequential decisions in a stochastic environment.

The significant new contribution of the work we present in this paper is to consider sequentiality not only in agents' actions but also in the allocation of resources.[1] In the single agent case, an agent might plan to change, in the midst of execution, how it utilizes its limited capacity. For example, it might return to the "toolbox" at its base station to drop off one instrument (e.g., a camera) and pick up another (e.g., a gripper). It might power down one peripheral and power up another, or terminate one process to create another. In the multiagent case, agents might at particular times swap who possesses or controls different instruments, or who holds the licenses for various software packages.

More precisely, in this paper we present formulations for defining, and algorithms for solving, several classes of single- and multi-agent *sequential* resource allocation decision problems for agents acting in stochastic environments. These problems are characterized by agents operating in multiple *phases*, where the set of resources held by an agent (and thus the actions that it can perform) are constant within a phase, but can change from one phase to another. The challenges that we tackle in this paper thus involve deciding how resource constraints should drive agents' overall missions to be best broken up into planned phases, and how agents should decide which resources to hold in each phase. As we shall see, these questions are intertwined with each other, and also with questions about how agents should formulate policies for pursuing their objectives in each phase.

### 1.1 Simple Illustrating Single-Agent Example

To drive home the problem in a very simple form, and to provide a running example that we will use to illustrate the formalisms, notations, and algorithms in the coming sections, we here present a simple schematic example to illustrate the Single-agent Resource-driven

1. This paper brings together and significantly extends work previously reported in conferences (Wu & Durfee, 2005, 2007a).

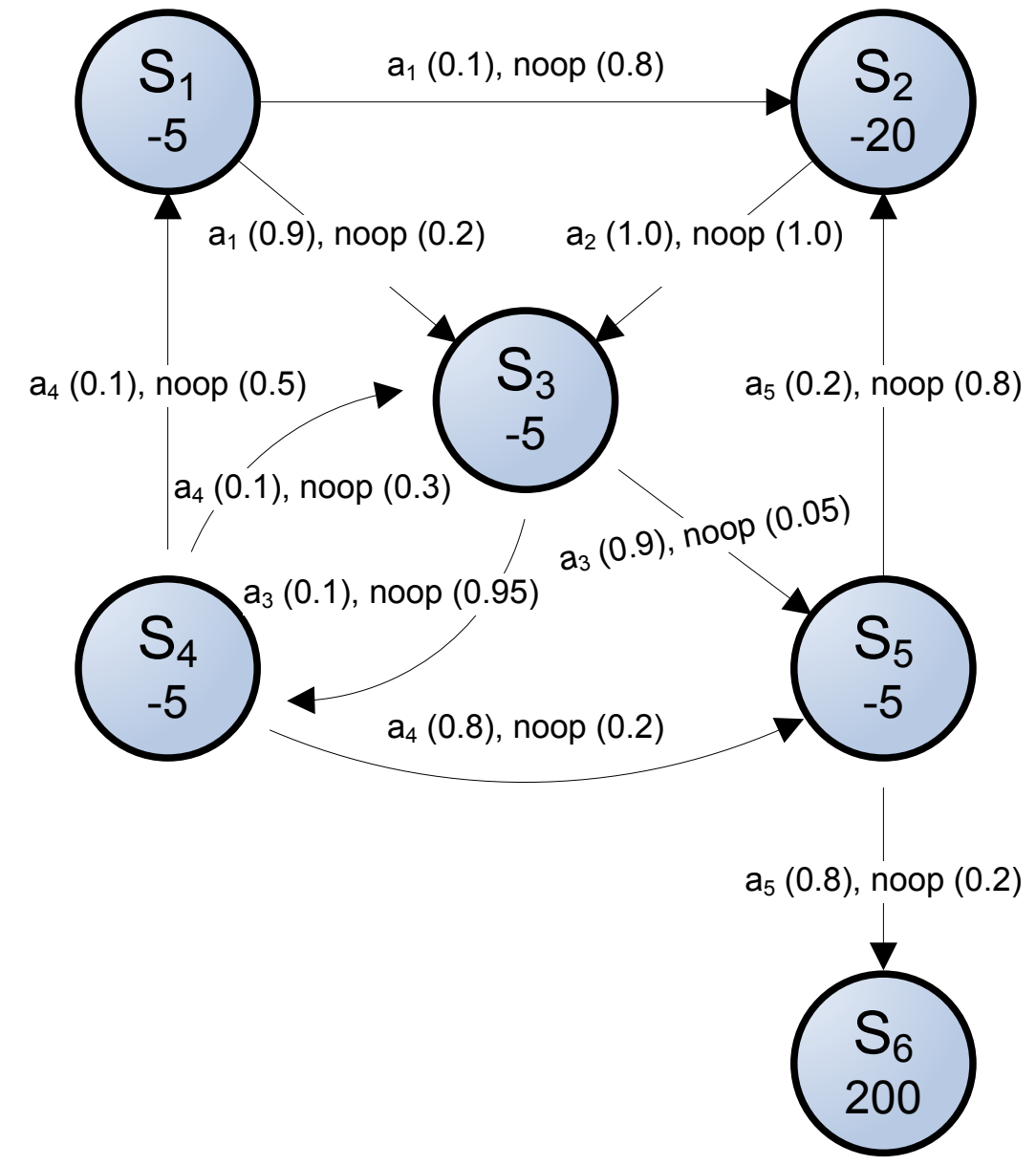

Figure 1: A simple single-agent example.

Mission Phasing (S-RMP) problem. We will present a simple multiagent example problem later in this paper.

In this problem, shown in Figure 1, the agent begins in state $S_1$ and moves among states $S_1$ through $S_5$ until it reaches and stops in state $S_6$. Associated with each state $S_i$ is the reward $r_i$ the agent receives for reaching that state, so, for example, reaching state $S_2$ is bad (incurring a reward of -20) while reaching state $S_6$ is good (providing a reward of +200).

The agent has some degree of control over its trajectory among states based on the action that it chooses to take in each state. For example, in state $S_1$ it has a choice of two actions, $a_1$ or *noop* (where *noop* is the "action" of not taking any action, passively letting the environment dynamics change the agent's state). As shown in the figure, if it takes action $a_1$ it has a probability of 0.1 of reaching state $S_2$, and of 0.9 of reaching state $S_3$, whereas *noop* reaches states $S_2$ and $S_3$ with probabilities 0.8 and 0.2 respectively. So, if we look at the agent's first decision, it would appear that it should choose action $a_1$ over *noop* to reduce the likelihood of the higher negative reward for reaching $S_2$.

However, to actually take action $a_1$, the agent needs to have a particular resource, which we will call $o_1$. More generally, for each of the actions $a_i$ that the agent can take, it needs resource $o_i$, other than for the *noop* action (which for notational convenience we will sometimes refer to as $a_0$), which does not require any resources. Thus, if when setting out from state $S_1$ the agent anticipated taking actions $a_1$, $a_3$, and $a_5$, say, it would want to set out having resources $o_1$, $o_3$, and $o_5$.

Unfortunately for the agent in this simple example, we will say that its capacity is limited such that it can have at most only a single resource at any given time. Now, before setting out from $S_1$ it will need to decide which resource would, in expectation, allow it to make action choices that would maximize its total reward by the time it reaches $S_6$. This is a simple instance of the type of problem solved by Dolgov and Durfee (2006), and we will show their solution shortly as a stepping stone to our algorithm.

In the S-RMP problem, we consider a generalization of Dolgov and Durfee's problem, where the agent has access to instances of all resources $o_i$ not only in state $S_1$, but also in some other states. After the agent sets off from state $S_1$ with the resource it considers most valuable, it traverses states until it reaches one of these other states, say state $S_i$. In state $S_i$, the agent can reconfigure its resources, subject as always to its capacity constraints, and thus switch to a new *phase* of execution, where the set of actions that it can take are different. We thus refer to states like $S_i$ (and, in a degenerate way, $S_1$) as *phase-switching states*. Referring back to our earlier examples of capacity-limited agents, a phase-switching state could correspond to the agent arriving at a location holding a cache of instruments (a "toolbox"), or it could correspond to reaching a time, place, or situation (e.g., a "holding pattern") where the environment is less dynamic such that the agent can safely power down some peripherals and power up others. Simple human examples of phase-switching states include the state I am in when I enter a parking lot, where now I can access my car, and the state of being on a highway entrance ramp, where as a driver I am given a buffer to change my driving behaviors and car speed to prepare to safely merge into high-speed traffic.

As we shall see, even in problems where the phase-switching states are static and predefined for the agent, phase-switching opportunities can complicate the agent's decisions about what actions to take, because now it might choose actions not only based on the rewards of reaching states but also based on the benefits to future action choices of reaching phase-switching states. Further, in some classes of problems, an agent itself might be able to decompose its problem into phases by deciding which states it would like as phase-switching states. For example, it could choose where it would like "toolboxes" (or entrance ramps) placed in its environment, or in which circumstances it would like to be buffered from environmental dynamics as it reconfigures itself. The agent would generally face constraints on creating phase-switching states, such as bounds on the number of such states (e.g., it might have a limited number of "toolboxes" to distribute), or incur a cost each time it creates a state in which it is buffered (e.g., every "holding pattern" introduces costly delays).

Thus, referring back to Figure 1, the Single-agent Resource-driven Mission Phasing (S-RMP) problem generally involves optimally deciding which states (besides the start state $S_1$) to designate as phase-switching states, optimally allocating resources in each of those states, and optimally choosing actions enabled by the resources during each phase. As we shall see later, the multiagent extension to this further requires that the multiple agents agree on their phase-switching states, since resource reallocation means potentially swapping (control over) different resources, and for each switch how best to (re)distribute their limited resources amongst themselves.

### 1.2 Paper Overview

While the idea of reconfiguring resources to improve agent performance is fairly straightforward, as the preceding example suggests it can be a challenging problem to reconfigure resources optimally. The primary goal of our study in this paper is to design computationally efficient algorithms to exactly solve this class of challenging problems. Toward this end, we develop a suite of algorithms that can formulate complex resource-driven mission-phasing problems into compact mathematical formulations. Thereafter, by simultaneously solving the problem decomposition (phase creation), resource (re)configuration, and policy formulation problems, these algorithms can fruitfully exploit problem structure, which often results in a significant reduction in computational cost.

This paper is organized as follows. Section 2 introduces background techniques. Section 3 starts with a relatively simple single-agent resource-driven mission-phasing problem where phase-switching states are known *a priori*. Exploiting such fixed phase-switching states, we can work out a particular, efficient algorithm. We then describe solution algorithms for solving general resource-driven mission-phasing problems, in which an agent needs to determine for itself where to reconfigure resources, how to reconfigure resources, and what are optimal executable policies subject to the (re)configured resources. Section 4 extends our resource-driven mission-phasing techniques presented in Section 3 to a class of multiagent systems for sequentially allocating resources among a group of cooperative agents. This section follows a similar progression as in Section 3, in terms of giving the agents increasing latitude in determining when to reallocate resources. Then, we contrast our work with related work in Section 5, and finally, Section 6 concludes the paper with a summary of this work, and a discussion of questions that remain open together with possible future research directions.

## 2. Background

We will formulate the single- and multi-agent resource-driven mission phasing problems using the well-established formalism of Markov Decision Processes (MDPs), with extensions to constrained MDPs. This section summarizes the relevant aspects of these previously-developed formalisms, and illustrates them using the example previously discussed in Section 1.1.

### 2.1 Markov Decision Processes

In general, a classical discrete-time, fully-observable Markov Decision Process with a finite state space and a finite action space can be defined as a four-tuple $\langle S, A, P, R\rangle$ (Puterman, 1994), where:

- ☐ $S$ is a finite state space, represented as a set of $n$ states $\{1, ...i, ...n\}$.
- ☐ $A$ is a finite action space. For a state $i \in S$, $A_i \subseteq A$ represents the set of actions that can be executed at the state $i$.
- ☐ $P = \{p_{i,a,j}\}$ represents state transition probability where $p_{i,a,j}$ is the probability that the agent reaches state $j$ if it executes action $a$ in state $i$.

  For any state $i$ and action $a$, $\sum_j p_{i,a,j}$ must be no greater than one. $\sum_j p_{i,a,j} = 1$ means that the agent will always stay in the system when executing action $a$ in state $i$, while $\sum_j p_{i,a,j} < 1$ means that there is some probability of the agent being out of the system (which can be equivalently interpreted as the agent entering a *sink* state where the agent would stay forever) when executing action $a$ in state $i$ (Kallenberg, 1983).
- ☐ $R = \{r_{i,a}\}$ is the (bounded) reward function where $r_{i,a}$ is the reward that the agent will receive if it executes action $a$ in state $i$.

**Running Example: MDP Encoding.** *The example introduced in Section 1.1 (Figure 1) is easily represented as a MDP:*

- ☐ $S = \{S_1, S_2, ...S_6\}$.
- ☐ $A = \{a_0 = noop, a_1, a_2, ...a_5\}$.
- ☐ $P = \{p_{S_1,a_0,S_2} = 0.8, p_{S_1,a_1,S_2} = 0.1, ...\}$.
- ☐ $R = \{r_{S_1,a_0} = -5, r_{S_1,a_1} = -5, ...\}$.

The Markov decision process is an extension of the well-known Markov chain. The main property of a MDP is that it possesses the Markov property (Bellman, 1957): if the current state of a MDP at time $t$ is known, transitions to a new state at time $t+1$ only depend on the current state and the action chosen at it, but are independent of the previous history of states.

In a MDP, the decision-making agent chooses its actions based upon its observation of the current state of the world, with the motivation of maximizing its aggregate reward. A deterministic stationary policy for a MDP is defined as a mapping from states to actions: $\pi : i \to a$ where $i \in S$ and $a \in A_i$. The objective of the decision-making agent is to find an optimal policy that maximizes some predefined cumulative function of rewards. Let $\{i_0, i_1, ..., i_t, ...\}$ and $\{a_0, a_1, ..., a_t, ...\}$ represent particular state and action sequences generated by following the policy $\pi$ starting in state $i_0$, and let $E[\,]$ denote the expectation function. Then a typical cumulative reward function of a non-discounted MDP can be defined as:

$$U(\pi) = E[\sum_{t=0}^{\infty} r_{i_t,a_t}]$$

Similarly, the cumulative reward function of a discounted MDP with the discount factor $\gamma$ can be defined as:[2]

$$U(\pi) = E[\sum_{t=0}^{\infty} (\gamma)^t \times r_{i_t,a_t}]$$

Although in general the mission-phasing techniques in this paper will also apply to discounted MDPs and other contracting MDPs (Kallenberg, 1983; Puterman, 1994; Sutton & Barto, 1998), we illustrate them in this paper using transient, non-discounted MDPs.[3] Non-discounted MDPs were described above. In a transient MDP (in which $\sum_j p_{i,a,j} < 1$ at some states), an agent will eventually leave the corresponding Markov chain, after running a policy for a finite number of steps (Kallenberg, 1983). In other words, given a finite state space, it is assumed that the agent visits any state only a finite number of times for any policy, which in turn means that the total expected reward function $U(\pi) = E[\sum_{t=0}^{\infty} r_{i_t,a_t}]$ is bounded even for a non-discounted MDP. The running example problem of Section 1.1 is an example of this kind of MDP, where state $S_6$ acts as a "sink" state ($\sum_j p_{6,a,j} = 0$).

### 2.2 Linear Programming

The value iteration and policy iteration algorithms are widely used in solving classical MDPs (Kallenberg, 1983; Puterman, 1994; Sutton & Barto, 1998). However, it is surprisingly hard to extend these algorithms to incorporate additional constraints without considerably increasing

2. In this paper, $(a)^b$ represents an exponent, while $a^b$ represents a superscript.
3. The transient MDP of interest in this work is a subclass of contracting MDPs.

the size of the state space and/or the action space of the MDP model. For that reason, a number of researchers have proposed and utilized an alternative solution approach, which is based upon mathematical programming (Altman, 1998; Feinberg, 2000; Dolgov & Durfee, 2006). A procedure for formulating an MDP into a linear program (whose solution yields an optimal policy maximizing the total expected reward) is described below. Our work extends this approach.

Let $x_{i,a}$, which is often called the *occupation measure* or *visitation frequency* (e.g., Dolgov & Durfee, 2006), denote the expected number of times action $a$ is executed in state $i$. Then the function $\sum_i \sum_a x_{i,a} \times r_{i,a}$ can be used to represent the total expected reward, and the problem of finding an optimal policy to the MDP is equivalent to solving the following linear program:

$$\max \sum_i \sum_a x_{i,a} \times r_{i,a} \tag{1}$$

subject to:

$$\sum_a x_{j,a} = \alpha_j + \sum_i \sum_a p_{i,a,j} \times x_{i,a} \qquad : \forall j$$

$$x_{i,a} \geq 0 \qquad : \forall i, \forall a$$

where $\alpha_j$ is the probability that the agent is initially in state $j$, and the constraint (named the *probability conservation* constraint) $\sum_a x_{j,a} = \alpha_j + \sum_i \sum_a p_{i,a,j} \times x_{i,a}$ guarantees that the expected number of times state $j$ is visited must equal the initial probability distribution at state $j$ plus the expected number of times state $j$ is entered via all possible transitions.

When the linear program Eq. 1 is solved, it is trivial to derive an optimal policy that specifies the action(s) to take in a given state. Specifically, a policy $\pi$ that assigns a probability of $\frac{x_{i,a}}{\sum_a x_{i,a}}$ to executing action $a$ in state $i$ will maximize the total expected reward. If any state-action probability in $\pi$ has a value other than zero or one, the optimal policy is randomized; otherwise it is deterministic.

**Running Example: MDP Policy Formulation.** *If we consider the problem introduced in Section 1.1 (Figure 1) but ignore the agent's capacity constraints (such that the agent has in every state the resources needed for its full set of actions $A = \{a_0, a_1, ...a_5\}$), then this corresponds to a classical (what we will refer to as an* unconstrained*) MDP. Using the above policy formulation algorithm, the agent can easily compute its optimal policy, which is $[S_1 \rightarrow a_1, S_2 \rightarrow noop/a_2, S_3 \rightarrow a_3, S_4 \rightarrow a_4, S_5 \rightarrow a_5, S_6 \rightarrow noop]$, and the total expected reward is 174.65.*

### 2.3 Constrained MDPs

Formulating unconstrained MDPs as linear programs makes it straightforward to take into account additional constraints, including the agent capacity constraints and resource constraints. Several of such constrained optimization problems have been investigated by Dolgov and Durfee (2006). In what follows, we summarize that work.

A constrained MDP that models agent capacity limitations can be represented as $\langle \mathcal{M}, \alpha, \mathcal{C} \rangle$, where:

- ☐ $\mathcal{M}$ is the classical MDP (Section 2.1), represented as $\langle S, A, P, R \rangle$.

- ☐ $\alpha = \{\alpha_i\}$ indicates the probability distribution over initial states.
- ☐ $\mathcal{C}$ is the agent capacity constraints, represented as $\langle O, C, U, \Gamma, \hat{\Gamma} \rangle$, where:
  - ⋄ $O = \{o\}$ is a finite set of indivisible non-consumable execution resources, e.g., $O = \{\textit{camera, spectrometer, gripper, etc.}\}$.
  - ⋄ $C = \{c\}$ is a finite set of capacities of the agent, e.g., $C = \{\textit{weight, space, etc.}\}$.
  - ⋄ $U = \{u_{o,a,i}\}$ represents resource requirements for executing actions, where $u_{o,a,i} \in \{0, 1\}$ indicates whether the agent requires resource $o$ to execute action $a$ in state $i$.[4] For example, $u_{o=camera,\, a=take_picture,\, i=any_state} = 1$ says that a prerequisite for taking a picture is having a camera.
  - ⋄ $\Gamma = \{\tau_{o,c}\}$ defines resource capacity costs, where $\tau_{o,c}$ is the amount of agent capacity $c$ required to hold one unit of resource $o$. For example, $\tau_{o=camera,\, c=weight} = 2$ and $\tau_{o=camera,\, c=space} = 1$ says that carrying a camera will consume two units of the carrying weight and one unit of the carrying space of the agent.
  - ⋄ $\hat{\Gamma} = \{\hat{\tau}_c\}$ specifies the limits of the agent capacities, e.g., $\hat{\tau}_{c=weight} = 4$ denotes a maximum weight of four units that an agent can carry.

**Running Example: Constraint Formulation.** *In the simple running example from Section 1.1, the agent constraint components are:*

- ☐ $O = \{o_1, o_2, ...o_5\}$.
- ☐ $C = \{hold\}$.
- ☐ $U = \{u_{o_i,a_i,s_i} = 1 : \forall 1 \leq i \leq 5\}$.
- ☐ $\Gamma = \{\tau_{o_i,c_{hold}} = 1 : \forall 1 \leq i \leq 5\}$.
- ☐ $\hat{\Gamma} = \{\hat{\tau}_{c_{hold}} = 1\}$.

The linear programming formulation (Eq. 1) paves the way for incorporating agent capacity constraints. Namely, the capacity limitations can be modeled by adding the following mathematical constraints (shown in Eq. 2) on occupation measures $x_{i,a}$ to the linear program in Eq. 1.

$$\sum_o \tau_{o,c} \times \Theta\Big( \sum_i \sum_a u_{o,a,i} \times x_{i,a} \Big) \leq \hat{\tau}_c \qquad : \forall c \tag{2}$$

where $\Theta(z)$ is a step function, defined as

$$\Theta(z) = \begin{cases} 1 & z > 0 \\ 0 & otherwise \end{cases}$$

The constraint indicates that, given the resource requirement parameter $u_{o,a,i} = 1$, the agent will have to employ $\tau_{o,c}$ amount of its capacity $c$ to hold resource $o$ if it decides to execute action $a$ in one or more states $i$ in its policy.

4. To simplify the presentation, it is assumed that the resource requirement is binary, which implies that an agent will not be interested in more than one unit of a particular resource, but most of results presented in this paper can be generalized to non-binary resource requirement cases without much difficulty.

Note that the $\Theta(z)$ function is a nonlinear function. In general, directly solving nonlinear constrained optimization problems is difficult. Fortunately, there is a simple way to transform the nonlinear constraint Eq. 2 into linear constraints through introducing some integer variables (Dolgov & Durfee, 2006). The reformulation of Equation 2 is depicted below.

$$\frac{\sum_i \sum_a u_{o,a,i} \times x_{i,a}}{X} \leq \Delta_o \qquad : \forall o$$
$$\sum_o \tau_{o,c} \times \Delta_o \leq \hat{\tau}_c \qquad : \forall c$$
$$\Delta_o \in \{0,1\} \qquad : \forall o$$

where $\Delta_o$, a binary integer in the set $\{0,1\}$, is introduced to indicate whether the agent uses its limited capacity to hold resource $o$. $X$ is a constant that is no less than $\sup \sum_i \sum_a x_{i,a}$, which is applied to guarantee that $\frac{\sum_i \sum_a u_{o,a,i} \times x_{i,a}}{X}$ never exceeds one (because $\sum_i \sum_a u_{o,a,i} \times x_{i,a} \leq \sum_i \sum_a x_{i,a} \leq \sup \sum_i \sum_a x_{i,a} \leq X$). One way to compute $X$ is to solve an unconstrained MDP:

$$X = \max \sum_i \sum_a x_{i,a} \tag{3}$$

subject to:

$$\sum_a x_{j,a} = \alpha_j + \sum_i \sum_a p_{i,a,j} \times x_{i,a} \qquad : \forall j$$
$$x_{i,a} \geq 0 \qquad : \forall i, \forall a$$

To summarize, the constrained MDP that models the agent's capacity limitations can be formulated into a mathematical program Eq. 4 (i.e., by putting Eq. 1 and the above integer linear constraints together), whose solution will yield an optimal capacity usage configuration and an optimal executable policy.

$$\max \sum_i \sum_a x_{i,a} \times r_{i,a} \tag{4}$$

subject to:

$$\sum_a x_{j,a} = \alpha_j + \sum_i \sum_a p_{i,a,j} \times x_{i,a} \qquad : \forall j$$
$$\frac{\sum_i \sum_a u_{o,a,i} \times x_{i,a}}{X} \leq \Delta_o \qquad : \forall o$$
$$\sum_o \tau_{o,c} \times \Delta_o \leq \hat{\tau}_c \qquad : \forall c$$
$$x_{i,a} \geq 0 \qquad : \forall i, \forall a$$
$$\Delta_o \in \{0,1\} \qquad : \forall o$$

In Eq. 4, $p_{i,a,j}$, $r_{i,a}$, $\alpha_j$, $u_{o,a,i}$, $\tau_{o,c}$, $\hat{\tau}_c$, and $X$ are constants, while $x_{i,a}$ are continuous variables and $\Delta_o$ are binary integer variables, which indicates that Eq. 4 is a mixed integer linear program (MILP).

Mixed integer linear programming is the discrete version of linear programming with an additional requirement that particular variables must be integers. Although MILPs are NP-hard in the number of integer variables, they can be solved by a variety of highly optimized algorithms and tools (Cook, Cunningham, Pulleyblank, & Schrijver, 1998; Wolsey, 1998). Recently, there has been substantial progress on using MILPs in automated planning (Earl & D'Andrea, 2005; Kautz & Walser, 2000; van Beek & Chen, 1999; Vossen, Ball, Lotem, & Nau, 1999). The automated resource-driven mission-phasing techniques (which are also NP-hard as is shown later) presented in this paper are based upon the MILP as well.

**Running Example: Constrained MDP Solution.** *In the simple running example from Section 1.1, the constraint permits an agent to hold only one resource (and thus to be capable of executing an action other than noop in only one state). Given the MDP and constraint parameters from this problem, and computing the constant $X = 70.24$ using Eq. 3, we apply a MILP solver such as* CPLEX *(www.ilog.com) to easily derive an optimal solution to the MILP:*

$$
\begin{aligned}
&[(x_{1,0}, x_{1,1}), (x_{2,0}, x_{2,2}), (x_{3,0}, x_{3,3}), (x_{4,0}, x_{4,4}), (x_{5,0}, x_{5,5}), x_{6,0}] \\
=&[(3.47, 0), (3.03, 0), (5.21, 0), (4.95, 0), (0, 1.25), 1] \\
&[\Delta_1, \Delta_2, \Delta_3, \Delta_4, \Delta_5] = [0, 0, 0, 0, 1]
\end{aligned}
$$

*That is, the optimal policy is $[S_1 \to noop, S_2 \to noop, S_3 \to noop, S_4 \to noop, S_5 \to a_5, S_6 \to noop]$, and the corresponding total expected reward is reduced to 65.02 (from 174.65 in the unconstrained case) due to the limitation on the agent capacity. This is the optimal policy for the constrained agent that uses a single policy throughout its entire mission. We will use this example as we go along to illustrate the degree to which our automated mission-phasing techniques can improve that expected reward.*

## 3. Resource Reconfiguration in Single-Agent Systems

We now turn to our new techniques and results that build on the work of others as summarized in the previous section. In particular, we extend the representations and techniques for solving constrained MDPs where resources are allocated prior to execution, to *sequential* constrained MDPs where resource allocations can change during execution when particular states are reached. As previously mentioned when we described the example problem (Section 1.1), we refer to the intervals during which an agent's resources cannot change as a *phase*, and the states that connect phases (representing an opportunity–but not an obligation–to change the resource allocation) as *phase-switching* states. We assume that the full complement of resources (e.g., a full "toolbox") is available at each phase-switching state, and that an agent cannot pick up or discard a resource except at a phase-switching state; relaxing this assumption is discussed as future work (Section 6.2).

At the extreme, if every state is a phase-switching state, then the agent is effectively unconstrained (unless there can exist an action whose necessary resources' total capacity requirements alone exceed the agent's capacity limits). In general, however, there will be restrictions on which states can (or should) be phase-switching states. We will consider several cases in this section. These range from where the phase-switching states are inherently predetermined by the environment (e.g., the placement of "toolboxes," or of "shelters" from domain dynamics, is dictated to the agent), to where the number of phase-switching states is bounded by the environment but which states are designated as phase-switching states is decided by the agent (e.g., the number of "toolboxes" or "shelters" is fixed but the agent can choose where they are placed), to where the number of phase-switching states is unbounded but where creating a phase-switching state incurs a cost (e.g., "toolboxes" or "shelters" can be bought and placed as the agent wishes) such that the agent will want to be selective so as to not spend more on creating a phase-switching state than the improvement the phase-switching state will make to its expected reward.

This section begins by giving a formal definition of the single-agent resource-driven mission-phasing (S-RMP) problem in Section 3.1, and then Section 3.2 analyzes and discusses the computational complexity of the S-RMP problem, illustrating why standard approaches are computationally intractable for solving the problem. Section 3.3 and Section 3.4 present, analyze, and illustrate solution algorithms for the variations of the S-RMP problem mentioned above. We present experimental results in Section 3.5, where the effectiveness and efficiency of our automated mission-phasing techniques are empirically evaluated. Finally, Section 3.6 summarizes the contributions of the work described in this section.

### 3.1 Problem Definition

Formally, a *single-agent resource-driven mission-phasing* (S-RMP) optimization problem is a generalization of the constrained MDP optimization problem presented in Section 2.3, composed of a Markov decision process $\mathcal{M}$, an initial probability distribution $\alpha$, agent capacity constraint $\mathcal{C}$, and resource reconfiguration constraint[5] $\mathcal{R}$, where:

- ☐ $\mathcal{M}$ is a classical MDP, as described in Section 2.1.
- ☐ $\alpha = \{\alpha_i\}$ is a probability distribution over states, where $\alpha_i$ is the probability that the agent starts in state $i$.
- ☐ $\mathcal{C}$ is the agent capacity constraint, as described in Section 2.3.
- ☐ $\mathcal{R}$ is the resource reconfiguration constraint (sometimes also called phase-switching constraint) that specifies restrictions on creating phase-switching states at which the constrained agent can reconfigure its resources and adjust its use of its limited capacities. A typical resource reconfiguration constraint $\mathcal{R}$ can be formulated as $\langle \lambda, \hat{\lambda} \rangle$ (and one of its generalizations will be discussed in Section 3.4.2), where:
  - ◇ $\lambda = \{\lambda_i\}$ indicates phase-switching state creation costs, where $\lambda_i$ denotes the cost for making state $i \in S$ into a phase-switching state.
  - ◇ $\hat{\lambda} \geq 0$ specifies a cost limit for creating phase-switching states.

For notational convenience, we also define $S_\epsilon \subseteq S$ as the set of *eligible* phase-switching states (indicating which of the states in $S$ can potentially become a phase-switching state). By definition, $S_\epsilon$ is a (not necessarily proper) subset of $S$, where for each $i \in S_\epsilon$, $\lambda_i \leq \hat{\lambda}$. That is, a state in $S$ is inherently *ineligible* to be a phase-switching state if its cost to make into a phase-switching state exceeds the agent's cost limit *all by itself*.

5. Because resource reconfiguration comes along with phase switching, in the following discussion, resource reconfiguration constraints are sometimes called phase-switching constraints to improve readability.

**Running Example: Resource Reconfiguration Constraints.** *Given the example in Section 1.1, here are a few of many possible specifications of the Resource Reconfiguration Constraints.*

- ☐ *When* $\lambda_1 \leq \hat{\lambda}$ *and* $\lambda_i > \hat{\lambda} : \forall i \neq 1$, *then* $S_\epsilon = \{S_1\}$ *and this is the constrained MDP situation described in Section 2.3.*

- ☐ *When* $\lambda_i = 0 : \forall i$ *(so,* $S_\epsilon = S$*), this is the case described at the beginning of this section where every state is a phase-switching state, and thus this equates to an unconstrained MDP unless the agent has at least one action that needs resources whose capacity requirements exceed a capacity constraint.*

- ☐ *If* $S_\epsilon \subset S$ *is predefined, then* $\lambda_i = 0 : \forall i \in S_\epsilon$ *but* $\lambda_i > \hat{\lambda} : \forall i \in S - S_\epsilon$. *This is the case where the phase-switching states are dictated to the agent.*

- ☐ *When* $\lambda_i = 1 : \forall i$, *and* $\hat{\lambda} = n$ *where* $0 < n < |S|$, *then* $S_\epsilon = S$ *and this is the case where the agent can select any subset of* $n$ *states to be phase-switching states.*

- ☐ *When* $\lambda_i > 0 : \forall i$, *and* $\hat{\lambda} = \infty$, *then* $S_\epsilon = S$ *and this is the case where the agent could turn any (and all) states into phase-switching states, but if the costs incurred are subtracted from the agent's reward it might choose to leave some (or most!) states as non-phase-switching states.*

Given the inputs $\mathcal{M}$, $\alpha$, $\mathcal{C}$, and $\mathcal{R}$, the objective of the S-RMP optimization problem is to maximize the total expected utility of the capacity-restricted agent by identifying a set of phase-switching states $S' = \{s^k\} \subseteq S_\epsilon$ that decompose the overall problem into a collection of phases, and, for each phase $k$, determining a resource configuration $\Delta^k$ and an executable policy $\pi^k$ that should be adopted by the agent at the entry to that phase (at state $s^k$).

Specifically, from a constrained optimization perspective, the S-RMP optimization problem can be formulated as follows:

Objective:

$$\text{maximize the utility of the overall policy } \pi$$

subject to the following constraints:

i) The set of phase-switching states $S' = \{s^k\}$ should satisfy the phase-switching constraint $\mathcal{R}$.

ii) Within each phase $k$, resource configuration $\Delta^k$ should satisfy the agent capacity constraint $\mathcal{C}$.

iii) Within each phase $k$, policy $\pi^k$ should be executable with respect to the resource configuration $\Delta^k$.

iv) The overall policy $\pi$ is composed of phase policies $\pi^k$, i.e., phase policy $\pi^k$ is adopted by the agent when it encounters a phase-switching state $s^k \in S'$ in the midst of its execution.

Clearly, the S-RMP optimization problem involves three intertwined components: i) problem decomposition, ii) resource configuration, and, iii) policy formulation. Problem decomposition (which creates phase-switching states) lays the foundation for resource configuration and reconfiguration; resource configuration dictates what policies are executable in each phase; policy formulation determines transitions within and among phases as well as what rewards can be accrued by the agent, which in turn determines the utility of problem decomposition and resource (re)configuration.

Each of these three component problems and some combinations of them have been investigated in a number of research fields (but none of the prior approaches is computationally tractable to the S-RMP optimization problem that tightly couples problem decomposition, resource configuration, and policy formulation). A comprehensive discussion that contrasts our work with prior work is postponed to Section 5 after our computationally efficient solution approach is presented.

### 3.2 Computational Complexity Analysis

Before describing our new solution techniques for S-RMP problems, we first analyze the computational complexity of the S-RMP optimization problem and illustrate why standard approaches are not computationally tractable for solving it.

**Theorem 3.1.** *S-RMP optimization is NP-complete.*

*Proof:* The proof of S-RMP optimization being NP-hard is trivial, because one of its special cases, which includes only one phase (i.e., the agent can only configure its resources at the beginning of mission execution), has been proven to be NP-hard through a reduction from the well-known KNAPSACK problem (Dolgov, 2006; Dolgov & Durfee, 2006).

The presence in NP can be proven in the following way. For a MDP with $n$ states, it is clear that there can be at most $n$ phases (i.e., $n$ phases in the extreme situation where every state is a phase-switching state). By featuring phase id (assuming each phase has a unique id) in the state representation, a generalized MDP with at most $n^2$ states can be constructed in polynomial time and the phase policies can be combined into an overall policy to this generalized MDP in polynomial time too. Given the generalized MDP and its policy, the problem is reduced to solving a Markov chain. Since a Markov chain can be verified in polynomial time, S-RMP optimization is in NP.

Given its presence in both NP and NP-hard, S-RMP optimization is proven to be NP-complete. □

S-RMP's complexity is evident when considering how straightforward solution methods would perform. In previous work (Wu, 2008), we compare our solution method to both a brute-force search algorithm and a MDP-expansion-based approach. We briefly summarize that comparison here. The brute-force search algorithm enumerates all possible problem decomposition schemes (legal combinations of phase-switching states), and, for each scheme enumerates all possible ways to configure and reconfigure resources, and, finally, for each possible problem decomposition and resource (re)configuration, derives optimal phase policies that are executable with respect to the configured resources. This quickly becomes intractable for non-trivial S-RMP optimization problems. The MDP-expansion-based approach instead incorporates resources into the state representation and models resource reconfiguration activities as explicit actions, and is more efficient than the brute-force algorithm, but still scales poorly because folding resources into states increases the state space size exponentially.

Neither of these approaches exploit key problem structure stemming from the coupled problems. The brute-force approach deals with S-RMP component problems in isolation and sequentially, while the MDP-based approach combines the resource-configuration and policy-formulation components in a naïve way that results in an exponentially larger policy formulation problem. In contrast, as we will see, our new solution algorithms can take advantage of problem structure by formulating problem decomposition, resource configuration, and policy formulation problems into a compact mathematical program and solving these component problems simultaneously, using a highly optimized tool. As will be shown in Section 3.5, the algorithms presented in this section can often find exact solutions to a complex S-RMP problem within a reasonable time.

### 3.3 Exploiting Fixed Phase-Switching States

So far, we have formally defined the S-RMP optimization problem and theoretically analyzed its computational complexity. In this and the next subsections, we present and illustrate our computationally efficient automated mission-phasing algorithms for solving S-RMP optimization problems.

We begin by first examining the simple variation of the S-RMP optimization problem where the phase-switching states are predetermined; that is, where the phase-switching states $S_\epsilon \subset S$ are given to the agent, such that $\forall i \in S_\epsilon$, $\lambda_i = 0$. This variation fits problems where the opportunities to reconfigure resources and switch policies (e.g., the placement of equipment and/or technicians needed for (un)loading or refitting) are dictated by the agent's environment rather than being a choice of the agent. Exploiting the fact that phase-switching states are fixed, we can devise a particular, efficient algorithm (while a more general but generally slower algorithm will be presented in Section 3.4).

Decomposition techniques for planning in stochastic domains are widely used for large environments with many states (and a detailed discussion of problem decomposition techniques will be given in Section 5). In those approaches, states are partitioned into small regions, a policy is computed for each region, and then these local policies are pieced together to obtain an overall policy (Parr, 1998; Precup & Sutton, 1998; Lane & Kaelbling, 2001). Our automated mission-phasing techniques are analogous to those decomposition techniques — partitioning a mission into multiple phases leads to smaller state and action spaces in each phase — though our motivation for mission phasing is to handle the constraints on policies agents can execute rather than to reduce the computational cost during policy formulation. Nonetheless, we can exploit these ideas.

Our algorithm for solving S-RMP optimization problems with predefined phase-switching states is based upon *abstract MDPs*. An abstract MDP is composed of abstract states, each of which corresponds to a mission phase. The "action" for an abstract state is the policy used in its corresponding phase (which is conceptually similar to *options*, Sutton, Precup, & Singh, 1999). It is here assumed that none of the constraints is associated with more than one phase. The discussion of more general constraints is postponed to the next subsection.

Since it is assumed that agent constraints in one phase cannot be affected by policy choices in another phase, the abstract MDP is an unconstrained MDP (at the abstract level) even though internally each phase is still a constrained MDP. The algorithm thus uses a policy iteration approach at the abstract level together with an embedded MILP solver within phases. The embedded MILP solver finds possible executable policies and their expected rewards for each of the phases, while different policies may have different probabilities of reaching the various phase-switching states at the "edges" of the phase. The outer policy iteration algorithm at the abstract level iteratively searches for the combination of phase policies that maximizes the reward across the whole mission.

The detailed procedure of the abstract MDP solver is illustrated below:

1. ***Partitioning the mission into phases.***
   When phase-switching states are given, partitioning a mission into multiple phases is straightforward. Start from a phase-switching state, and then keep expanding through all connected transitions until encountering other phase-switching states. The resulting state space is the phase state space corresponding to that phase-switching state.[6]

2. ***Policy iteration.***
   The following policy iteration algorithm is adopted after the state space is partitioned.

   (a) For each phase-switching state $s$, solve the MDP corresponding to the phase beginning in that state as if it were an unconstrained MDP, and compute state value $V(s)$. These $V(s)$ are used as initial values of phase-switching states since they provide informed (as opposed to random) estimates that in our experience tend to work well, especially for under-constrained problems.

   (b) In the abstract MDP, each phase is treated as an abstract state and each policy for a phase is treated as an abstract action for that phase's abstract state. The policy iteration algorithm alternates between the following two steps:

   *Policy improvement*: Rather than enumerating all possible policies (abstract actions) for a phase (abstract state), the algorithm uses a constrained MDP solver (that was shown in Eq. 4) to calculate the optimal policy in the phase, given the current values $V(s)$ associated with the (outgoing) neighboring phase-switching states.

   *Policy evaluation*: Given abstract actions, calculate $V(s)$ for each phase-switching state $s$. For small state spaces, standard linear algebra methods are often the best solutions for policy evaluation. For larger state spaces, a simplified value iteration algorithm might be preferable (simplified because the policy in each phase is fixed) (Puterman, 1994).

Unlike much "best-response" hill-climbing work, the above abstract MDP has fixed state transition functions and fixed reward functions in both the abstract level and the phase level because the agent enters a phase always at the same phase-switching state, which guarantees the above policy iteration algorithm will return an optimal solution.

**Theorem 3.2.** *The abstract MDP policy iteration procedure will converge to an optimal solution.*

6. Note that exploiting factored state representations can lead to other, more efficient (although generally only approximate) algorithms for partitioning (Kim & Dean, 2001; Guestrin, Koller, Parr, & Venkataraman, 2003).

*Proof:* In each iteration, the new abstract policy is a strict improvement over the previous one. Since the total expected reward of the abstract MDP is bounded (because the total expected reward of the corresponding unconstrained MDP is bounded), the iteration procedure will eventually converge.

At the convergence point, both the phase MDPs and the abstract MDP satisfy the Bellman optimality equation (because of the nature of the linear programming solver and the policy iteration algorithm), indicating that the derived policy is an optimal policy. □

**Running Example: Optimizing for Predetermined Phase-Switching States.** *We now return to our running example introduced in Section 1.1 to illustrate how the total expected reward can be improved if the agent can reconfigure its resources at some states. Let us say that the agent is told that it is able to reconfigure resources and switch policies at states $S_1$, $S_3$ and $S_4$. These three phase-switching states decompose the example problem into three phases. The corresponding abstract MDP is constructed and shown in Figure 2, which is composed of three abstract states.*

*Using the abstract MDP policy-iteration algorithm just described and using the same parameters as before (especially that an executable policy cannot have more than one action that is not a noop), the state values of the phase-switching states eventually converge to*

$$V(S_1) = 113.65 \qquad V(S_3) = 120.65 \qquad V(S_4) = 123.05$$

*The optimal policy in phase I is $[S_1 \rightarrow a_1, S_2 \rightarrow noop]$ (with resource $o_1$), the optimal policy in phase II is $[S_2 \rightarrow noop, S_3 \rightarrow noop, S_5 \rightarrow a_5, S_6 \rightarrow noop]$ (with resource $o_5$), and the optimal policy in phase III is $[S_2 \rightarrow noop, S_4 \rightarrow noop, S_5 \rightarrow a_5, S_6 \rightarrow noop]$ (also with resource $o_5$). The total expected reward is now 113.65, which is 74.8% higher than the expected reward for the constrained MDP case (where resource allocation only happens in state $S_1$) thanks to the additional phase-switching states.*

Thanks to the policy iteration procedure, the abstract MDP solver generally converges quickly. However, it should be noted that two limitations are inherent in the abstract MDP solver. One of the limitations is that the abstract MDP solver requires that phase-switching states are known *a priori*, which restricts its applicability (although we can combine it with some phase-switching-state heuristic search techniques). The other limitation is due to the possible existence of constraints running across multiple phases. The abstract MDP solver cannot cope with constraints associated with multiple abstract states, such as restrictions on the expected number of visits to a particular state that belongs to multiple phases. In contrast, the general S-RMP solution algorithms that we present in the next section do not have such limitations.

### 3.4 Determining Optimal Phase-Switching States

In a general S-RMP optimization problem, phase-switching states are not completely predetermined. Instead, given a defined set of eligible phase-switching states $S_\epsilon$, a set of costs $\{\lambda_i\}$ (where $\lambda_i$ denotes the cost for making eligible state $i$ into a phase-switching state), and a cost limit $\hat{\lambda}$, the objective of the agent is to find an optimal phase-switching set $S' \subseteq S_\epsilon$ subject to $\sum_{i \in S'} \lambda_i \leq \hat{\lambda}$, along with optimal resource configurations and optimal executable policies within each phase, to maximize its expected cumulative reward.

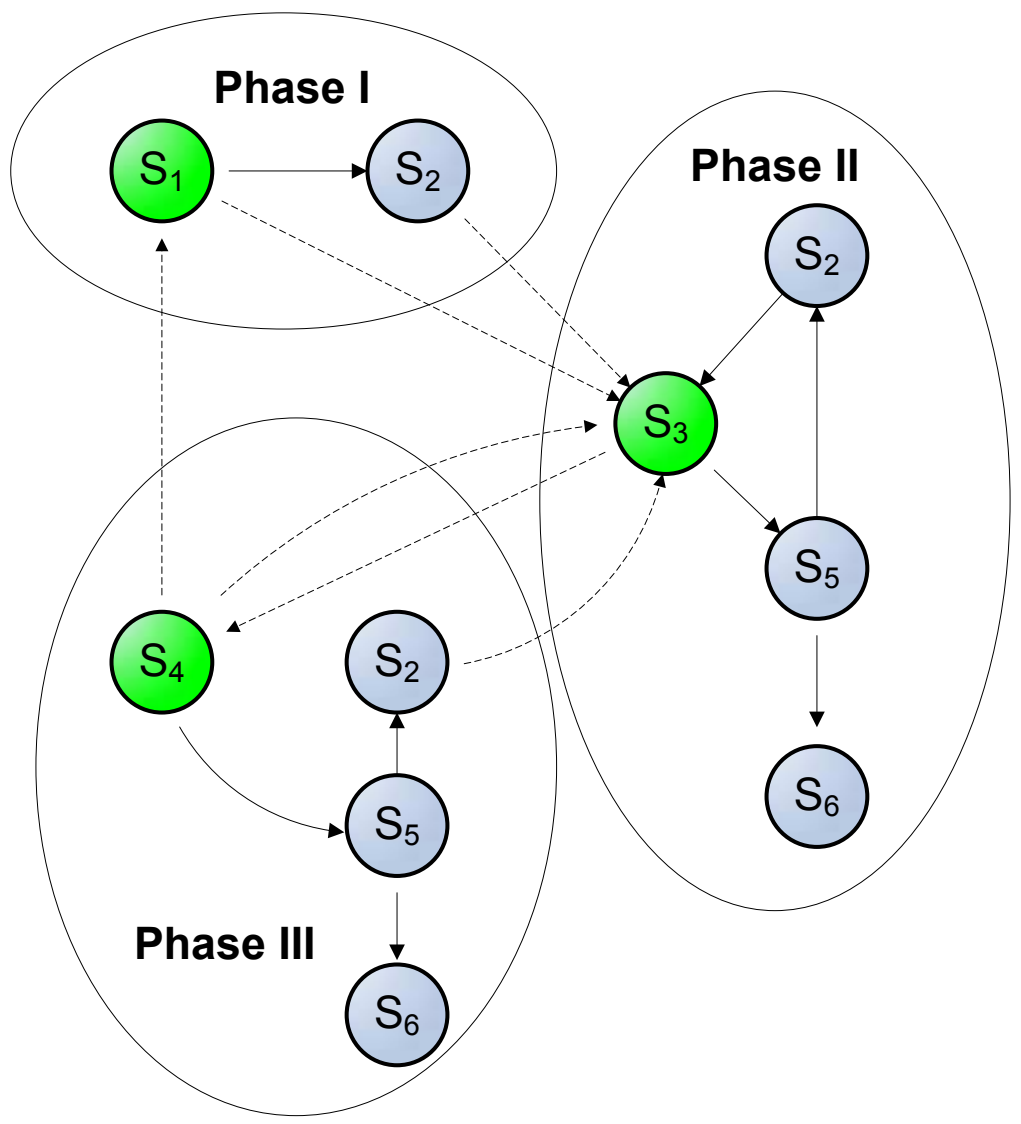

Figure 2: An abstract MDP with three phases.

As mentioned, the abstract MDP solver presented in Section 3.3 cannot be directly used for the general S-RMP optimization problem. In this section, we construct a mixed integer linear program, the solution to which yields the optimal set of phase-switching states maximizing the total expected reward, as well as optimal resource configurations and executable policies within each phase. We make a simplifying assumption that a state with a positive probability $\alpha_j$ of being the initial state is always a phase-switching state. This assumption makes the presentation clearer and the representation more concise, as well as sidestepping the question of what the "default" agent policy might be (since that is what it would use if it could not configure resources in an initial state).

Let $x^k_{i,a}$ be the expected number of times action $a$ is executed in state $i$ within phase $k$. Clearly, if state $i$ is not reachable in phase $k$, then $x^k_{i,a} = 0$. Let $\alpha^k_j = \sum_a x^k_{j,a} - \sum_i \sum_a p_{i,a,j} \times x^k_{i,a}$ where $p_{i,a,j}$ is the state transition probability, then $\alpha^k_j$ provides a way to characterize transitions among phases. If state $j$ is not a phase-switching state, then $\alpha^k_j = 0$ for any $k$, since within any phase the expected number of times of visiting state $j$ ($\sum_a x^k_{j,a}$) must equal the expected number of times of entering state $j$ through all possible transitions ($\sum_i \sum_a p_{i,a,j} \times x^k_{i,a}$). If state $j$ is a phase-switching state, $\sum_k \alpha^k_j = \alpha_j$. Recall that $\alpha_j$ is the initial probability distribution for state $j$. $\sum_k \alpha^k_j = \alpha_j$ guarantees that the total expected number of times of visiting state $j$ must equal the initial probability distribution for state $j$ plus the total expected number of times of entering state $j$ through all possible transitions.

Now, we can formulate the S-RMP optimization problem into a mixed integer linear program, which is shown in Eq. 5. The formulation builds on the MILP for the constrained MDP given in Eq. 4, but incorporates phase information. The objective function $\sum_i \sum_a \sum_k x^k_{i,a} \times r_{i,a}$ in the MILP represents the total expected reward accumulated across all phases, where $r_{i,a}$ is the MDP reward function.

$$\max \sum_k \sum_i \sum_a x^k_{i,a} \times r_{i,a} \tag{5}$$

subject to:

*probability conservation constraints:*

$$\sum_a x^k_{j,a} = \alpha^k_j + \sum_i \sum_a p_{i,a,j} \times x^k_{i,a} \quad : \forall k, \forall j$$

$$\sum_k \alpha^k_j = \alpha_j \quad : \forall j$$

$$x^k_{i,a} \geq 0 \quad : \forall k, \forall i, \forall a$$

*capacity constraints:*

$$\frac{\sum_i \sum_a u_{o,a,i} \times x^k_{i,a}}{X} \leq \Delta^k_o \quad : \forall o, \forall k$$

$$\sum_o \tau_{o,c} \times \Delta^k_o \leq \hat{\tau}_c \quad : \forall c, \forall k$$

$$\Delta^k_o \in \{0,1\} \quad : \forall o, \forall k$$

*phase-switching constraints:*

$$\frac{\alpha^k_j}{X} \leq \Lambda_j \quad : \forall k, \forall j$$

$$\sum_j \lambda_j \times \Lambda_j \leq \hat{\lambda}$$

$$\Lambda_j \in \{0,1\} \quad : \forall j$$

- As stated above, the constraint $\sum_a x^k_{j,a} = \alpha^k_j + \sum_i \sum_a p_{i,a,j} \times x^k_{i,a}$ models the conservation of probability within each phase.

- The constraint $\sum_k \alpha^k_j = \alpha_j$ indicates the probability conservation constraint across phases, i.e., $\sum_k \alpha^k_j = \sum_k (\sum_a x^k_{j,a} - \sum_i \sum_a p_{i,a,j} \times x^k_{i,a}) = \sum_a x_{j,a} - \sum_i \sum_a p_{i,a,j} \times x_{i,a} = \alpha_j$, where $x_{i,a} = \sum_k x^k_{i,a}$ is the total expected number of times action $a$ is executed in state $i$.

- The capacity constraints $\frac{\sum_i \sum_a u_{o,a,i} \times x^k_{i,a}}{X} \leq \Delta^k_o$ and $\sum_o \tau_{o,c} \times \Delta^k_o \leq \hat{\tau}_c$ are a multi-phase version of the capacity constraints discussed in Eq. 4, where $X = \max \sum_i \sum_a x_{i,a}$ can be computed by using Eq. 3.

- $\Lambda_j$ in the constraint $\frac{\alpha_j^k}{X} \le \Lambda_j$ is a binary variable, where $\Lambda_j = 1$ when state $j$ is a phase-switching state, and $\Lambda_j = 0$ otherwise. We can prove $X \ge \sup \alpha_j^k$ as follows:

$$\begin{aligned}\sup \alpha_j^k &= \sup(\sum_a x_{ja}^k - \sum_i \sum_a p_{ij}^a x_{ia}^k) \\ &\le \sup \sum_a x_{ja}^k \\ &\le \sum_i \sum_a \sum_k x_{ia}^k \\ &= X\end{aligned}$$

  Therefore, this constraint and the constraint $\Lambda_j \in \{0,1\}$ guarantee that $\exists k : \frac{\alpha_j^k}{X} > 0 \Rightarrow \Lambda_j = 1$, which implies that a state must be a phase-switching state if there is some "transition leakage" at that state in any phase.

- The constraint $\sum_j \lambda_j \times \Lambda_j \le \hat{\lambda}$ says that the cost of creating phase-switching states must be no greater than the cost limit $\hat{\lambda}$.

- Other constraints denote the ranges of variables. Note that there are no range restrictions for $\alpha_j^k$.

By definition, $\Lambda_j$ and $\Delta_o^k$ in the solution to Eq. 5 indicate phase-switching states and resource configuration (within each phase), respectively. Once we have solved Eq. 5, we can use the resulting values of $x_{i,a}^k$ to derive an optimal overall policy as follows.

1. ***Compute the optimal policy for each phase.***
   This is the same as before — at state $i$, action $a$ is executed with probability $\pi_{i,a}^k = \frac{x_{i,a}^k}{\sum_a x_{i,a}^k}$. In the following discussion, we use $\pi^k = \{\pi_{i,a}^k\}$ to denote the phase policy in phase $k$.

2. ***Determine the phase policy to adopt at a phase-switching state.***
   This is also trivial. The agent should choose phase policy $\pi^k$ with probability $\frac{x_i^k}{\sum_k x_i^k}$ at phase-switching state $i$ for maximizing its total expected reward, where $x_i^k = \sum_a x_{i,a}^k$.

**Running Example: Selecting an Optimal Fixed Number of Phase-Switching States.** *We now illustrate the solution algorithm on our running example depicted in Figure 1. Recall that, as was shown in Section 3.3, when the agent is allowed to reconfigure its resources and switch its policy at $S_1$, $S_3$ and $S_4$, its total expected reward is 113.65 (higher than the reward 65.02 in one-shot constrained MDP case, but still much lower than the optimal reward 174.65 in the unconstrained MDP case). Rather than starting with predefined phase-switching states, we now assume that $\lambda_1 = 0$, $\lambda_{i \in \{2,\dots,6\}} = 1$, and $\hat{\lambda} = 2$. That is to say, two additional phase-switching states besides the initial state $S_1$ can be chosen by the agent from any states in the system.*

*We use the same transition probability $p_{i,a,j}$, reward $r_{i,a}$, initial probability distribution $\alpha_j$, resource requirement cost $u_{o,a,i}$, capacity cost $\tau_{o,c}$, capacity limit $\hat{\tau}_c$, and the constant value*

*$X$ as in Section 1.1. The phase-switching costs $\lambda_i$ and the cost limit $\hat{\lambda}$ are given above. The optimal integer solution to the mixed integer linear program Eq. 5 is:*

$$[\Lambda_1, \Lambda_2, \Lambda_3, \Lambda_4, \Lambda_5, \Lambda_6] = [1, 0, 1, 0, 1, 0]$$
$$\begin{vmatrix} \Delta_1^1, \Delta_2^1, \Delta_3^1, \Delta_4^1, \Delta_5^1 \\ \Delta_1^2, \Delta_2^2, \Delta_3^2, \Delta_4^2, \Delta_5^2 \\ \Delta_1^3, \Delta_2^3, \Delta_3^3, \Delta_4^3, \Delta_5^3 \end{vmatrix} = \begin{vmatrix} 1, 0, 0, 0, 0 \\ 0, 0, 1, 0, 0 \\ 0, 0, 0, 0, 1 \end{vmatrix}$$

*That is, the optimal set of phase-switching states is $S' = \{S_1, S_3, S_5\}$. Examining the continuous variables $x_{i,a}^k$ (not shown here because there are too many of them) shows that the total expected reward of the agent is 173.80, which is close to the optimal unconstrained reward of 174.65. The derived solution is to choose resource $o_1$ and adopt the policy $[S_1 \to a_1, S_2 \to noop]$ at $S_1$, switch to resource $o_3$ and policy $[S_3 \to a_3, S_4 \to noop]$ at $S_3$, and switch to resource $o_5$ and policy $[S_2 \to noop, S_5 \to a_5, S_6 \to noop]$ at $S_5$.*

#### 3.4.1 Variation: Maximizing the Total Reward, Accounting for Cost

This subsection demonstrates the extensibility of our MILP-based algorithm by showing how easily it can be revised to work for another useful variation of the S-RMP optimization problem where the phase-switching states are not predetermined (Section 3.3) nor is the cost of creating phase-switching states bounded (Section 3.4). We now assume that any state could be a phase-switching state, that as many states as desired could be phase-switching states, and that (similarly as in Section 3.4) there is a cost associated with treating a state as a phase-switching state. However, instead of being subject to some cost limits, these costs are now calibrated with the rewards associated with executing policies. Now the optimization problem is to maximize the total expected reward, accounting for the costs of creating phase-switching states, without predetermining which are the phase-switching states or how many there will be. As shown below, designing an algorithm for such problems is trivial. It is just a simple mathematical reformulation of Eq. 5. The details are presented in Eq. 6 (which omits the probability conservation and capacity constraints because they are unchanged from Eq. 5).

$$\max \sum_k \sum_i \sum_a x_{i,a}^k \times r_{i,a} - \sum_i \lambda_i \times \Lambda_i \tag{6}$$

subject to:

*probability conservation constraints (unchanged)*

*capacity constraints (unchanged)*

*phase-switching constraints:*

$$\frac{\alpha_j^k}{X} \leq \Lambda_j \qquad : \forall k, \forall j$$
$$\Lambda_j \in \{0, 1\} \qquad : \forall j$$

where $\lambda_i$ represents the cost for creating phase-switching state $i$, and the objective function $\sum_k \sum_i \sum_a x_{i,a}^k \times r_{i,a} - \sum_i \lambda_i \times \Lambda_i$ represents the total expected reward of the policy minus the cost for creating phase-switching states.

| Case | Phase-Switching States | Expected Utility |
|---|---|---|
| unconstrained (Section 2.1) | $\{s_1, s_2, s_3, s_4, s_5, s_6\}$ | -75.35 |
| one-shot (Section 2.3) | $\{s_1\}$ | 65.02 |
| 3 fixed phases (Section 3.3) | $\{s_1, s_3, s_4\}$ | 13.65 |
| 2 added chosen phases (Section 3.4) | $\{s_1, s_3, s_5\}$ | 73.80 |
| unlimited phases balancing cost (Section 3.4.1) | $\{s_1, s_5\}$ | 102.55 |

Table 1: Comparison of the solutions to the example problem, given that the cost of creating each additional phase-switching state is 50.

**Running Example: Selecting Optimal Phase-Switching States Based on Cost.** *Let us revisit our running example to illustrate how the above algorithm can be used to solve this variation of the S-RMP optimization problem. Suppose that* $\lambda_1 = 0$ *(assuming the initial state is already a phase-switching state) and* $\lambda_i = c$ *for any other state. Using the above MILP formulation (Eq. 6), we can find that when* $0 < c \leq 0.85$ *the optimal phase-switching states are* $[S_1, S_3, S_4, S_5]$*, when* $0.85 < c \leq 21.25$*, the optimal phase-switching states are* $[S_1, S_3, S_5]$*, when* $21.25 < c \leq 87.53$*, the optimal phase-switching states are* $[S_1, S_5]$*, and when* $c > 87.53$ *the optimal decision is not to create additional phase-switching states besides the initial state* $S_1$*. As expected, the number of phase-switching states decreases as the cost of creating phase-switching states increases.*

*As a specific example, when* $c = 50$*, the optimal set of phase-switching states is* $\{S_1, S_5\}$*. The optimal resource configuration and executable policy in the phase initiated at* $S_1$ *are* $\{o_3\}$ *and* $[S_1 \rightarrow noop, S_2 \rightarrow noop, S_3 \rightarrow a_3, S_4 \rightarrow noop]$*, respectively; the optimal resource configuration and executable policy in the phase initiated at* $S_5$ *are* $\{o_5\}$ *and* $[S_2 \rightarrow noop, S_3 \rightarrow noop, S_4 \rightarrow noop, S_5 \rightarrow a_5]$*, respectively. The policy utility is* $152.55\ (reward) - 50 \times 1\ (cost) = 102.55$*.*

*To better understand and compare this solution with the example solutions derived in the previous sections, Table 1 shows their solution utilities (where the utility is defined as the expected reward of the policy minus the cost of creating phase-switching states). Not surprisingly, the utility of the solution presented in this subsection is higher than the others since it is derived by the algorithm (Eq. 6) that explicitly balances the costs and expected benefits for creating phase-switching states.*

#### 3.4.2 Variation: Cost Associated with State Features

A final variation that we briefly describe, which has similarities to the multiagent approach described later in Section 4, is the case where the conditions that enable resource reconfiguration (phase switching) are associated with a subset of the world features, rather than a fully grounded state. As a simple example, consider the situation where resources (e.g., software packages, control over a satellite, etc.) can be licensed/leased at particular times and for particular intervals, such as hourly. That is, the agent will identify, at the start of each hour, which resources (within its capacity constraints) to hold for the next hour. It could be in any number of states (e.g., physical locations, pending task queue, etc.), but the resource reconfiguration can take place in any of them. Similarly, in the example where a robot reconfigures resources at a "toolbox", the critical feature for the phase switch is that it is in a state whose location feature corresponds to a toolbox, regardless of other state features (e.g., direction it is facing or battery level).

More generally, let us say that the MDP state space $S$ consists of $L$ disjoint subsets $\mathcal{S}_1$, $\mathcal{S}_2$, ..., $\mathcal{S}_l$, ..., $\mathcal{S}_L$, where each subset contains states that have identical values for the phase-switching feature(s) (e.g., all states in $S_l$ have clock time $t$). Thus, if any state within $\mathcal{S}_l$ is a phase-switching state then all states in $\mathcal{S}_l$ are phase-switching states as well. Let $\lambda_l$ denote the cost for making *all* the $\mathcal{S}_l$ states into phase-switching states, i.e., the cost to enable phase switching in worlds where the critical feature(s) take on their common value(s) in $\mathcal{S}_l$, and let $\hat{\lambda}$ denote the cost limit for creating phase-switching states. Clearly, this is a generalization of the previous phase-switching constraint: when every $\mathcal{S}_l$ contains exactly one state, this representation is equivalent to the phase-switching constraint $\mathcal{R}$ previously presented in Section 3.1.

The new mixed integer linear program with the generalized phase-switching constraint is formulated in Eq. 7, which again is very similar to Eq. 5, except for some minor revisions in the portion of phase-switching constraints. As before, constraints unchanged from Eq. 5 are not repeated.

$$\max \sum_k \sum_i \sum_a x^k_{i,a} \times r_{i,a} \tag{7}$$

subject to:

*probability conservation constraints (unchanged)*

*capacity constraints (unchanged)*

*phase-switching constraints:*

$$\frac{\alpha^k_j}{X} \le \Lambda_l \qquad : \forall k, \forall l, \forall j \in \mathcal{S}_l$$

$$\sum_l \lambda_l \times \Lambda_l \le \hat{\lambda}$$

$$\Lambda_l \in \{0,1\} \qquad : \forall l$$

where binary variable $\Lambda_l$ denotes whether $\mathcal{S}_l$ is a phase-switching set.

**Running Example: Selecting Optimal Phase Switching Features.** *In our running example, suppose now that the state space is composed of $\mathcal{S}_{l=1} = \{S_1\}$, $\mathcal{S}_{l=2} = \{S_2, S_3\}$, $\mathcal{S}_{l=3} = \{S_4, S_5\}$, and $\mathcal{S}_{l=4} = \{S_6\}$, and that $\lambda_{l=1} = 0$, $\lambda_{l\neq 1} = 1$, and $\hat{\lambda} = 1$. The solution to Eq. 7 will yield a policy with a reward* 165.68 *using phase-switching states $\{S_1, S_4, S_5\}$, where the spending of one unit of cost creates both phase-switching state $S_4$ and phase-switching state $S_5$.*

### 3.5 Experimental Evaluation

To this point, we have described variations of single-agent resource-driven mission-phasing problems and techniques for solving them, using a simple example to illustrate these ideas. Ultimately, the significance of these techniques hinges on their computational efficiency in solving problems that are more difficult. In this section, we give an empirical evaluation of our techniques focusing on problems in a more complex state space and with a larger resource set. Our experiments are implemented on a simplified Mars rover domain simulation in which an autonomous rover operates in a stochastic environment. Following much of the literature on similar problems (Bererton, Gordon, & Thrun, 2003; Dolgov & Durfee, 2006), the Mars rover domain is represented using a grid world.

### 3.5.1 Experimental Setup

The grid world (see Figure 3c) has some number of *wall* locations through which the rover cannot move. Each of the other locations is associated with an execution resource, which, if held by the rover, allows the rover to move with confidence and safety in that location to its desired next location. For example, the resources could be sensors for the conditions in different locations (dusty, foggy, overgrown, etc.), or different kinds of wheels (for navigating sand, rocks, etc.). The rover agent can also move without holding the appropriate resource in a location, but this will result in greater uncertainty in action outcomes (it could blunder or slip, and thus arrive in a different location than desired) and possibly cause damage to the rover, as detailed shortly.

In addition, there are multiple tasks randomly distributed in the grid world. When the rover reaches a location that has a task, if the rover currently carries the task-required execution resource, the rover can choose to perform a *do* action (that carries out the task) and receive a reward. Once any task is carried out, the mission is accomplished and the rover will leave the system (the experimental run terminates).

Our experiences running a variety of experiments indicate that the trends in the results presented in this section are not sensitive to exact parameter settings, but for the sake of reproducibility, we describe the detailed parameters below. The procedure of building a random grid world is illustrated in Figure 3. When a $n \times n$ grid world is built, 40% of the locations are randomly chosen as wall locations, and 10% of the locations are randomly chosen as task locations. To avoid simple test problems, we only use grid worlds whose number of reachable locations (from the rover's starting location) is greater than half of the total number of locations (i.e., greater than $n^2/2$).

At each task location, there is a task that could be accomplished by the rover and generate a reward. To make the problem interesting and challenging, we distinguish tasks by setting different rewards for them. We sort tasks by their Manhattan distances to the starting location of the rover (the smallest distance first), and let the $i^{th}$ task have a reward $i$. Therefore, it is not always true that the rover would desire and pursue high-reward tasks because low-reward tasks are closer to the rover and might be easier and safer to complete.

The rover always starts at the S (START) location in the left bottom corner of the grid (which is never assigned to be a wall), and its objective is to maximize its expected reward. At each time step, the rover chooses an action in its action set $\{$*wait*, *up*, *left*, *down*, *right*, *safe-up*, *safe-left*, *safe-down*, *safe-right*, *do*$\}$. Actions *wait*, *up*, *left*, *down*, and *right* can be executed without requiring the rover to carry any particular resource. In contrast, performing a safe-moving action *safe-up*, *safe-left*, *safe-down*, or *safe-right* in a non-wall location requires a particular resource (related to that location), which is randomly uniformly selected from resource set $O$ when the problem is built. Analogously, performing action *do* at a task location requires a particular resource that is also randomly uniformly selected. It should be pointed out that performing an action needs at most one resource, but a resource may (by chance) enable the agent to move safely in multiple locations, and/or carry out multiple tasks. The resource requirement information is known to the rover *a priori*.

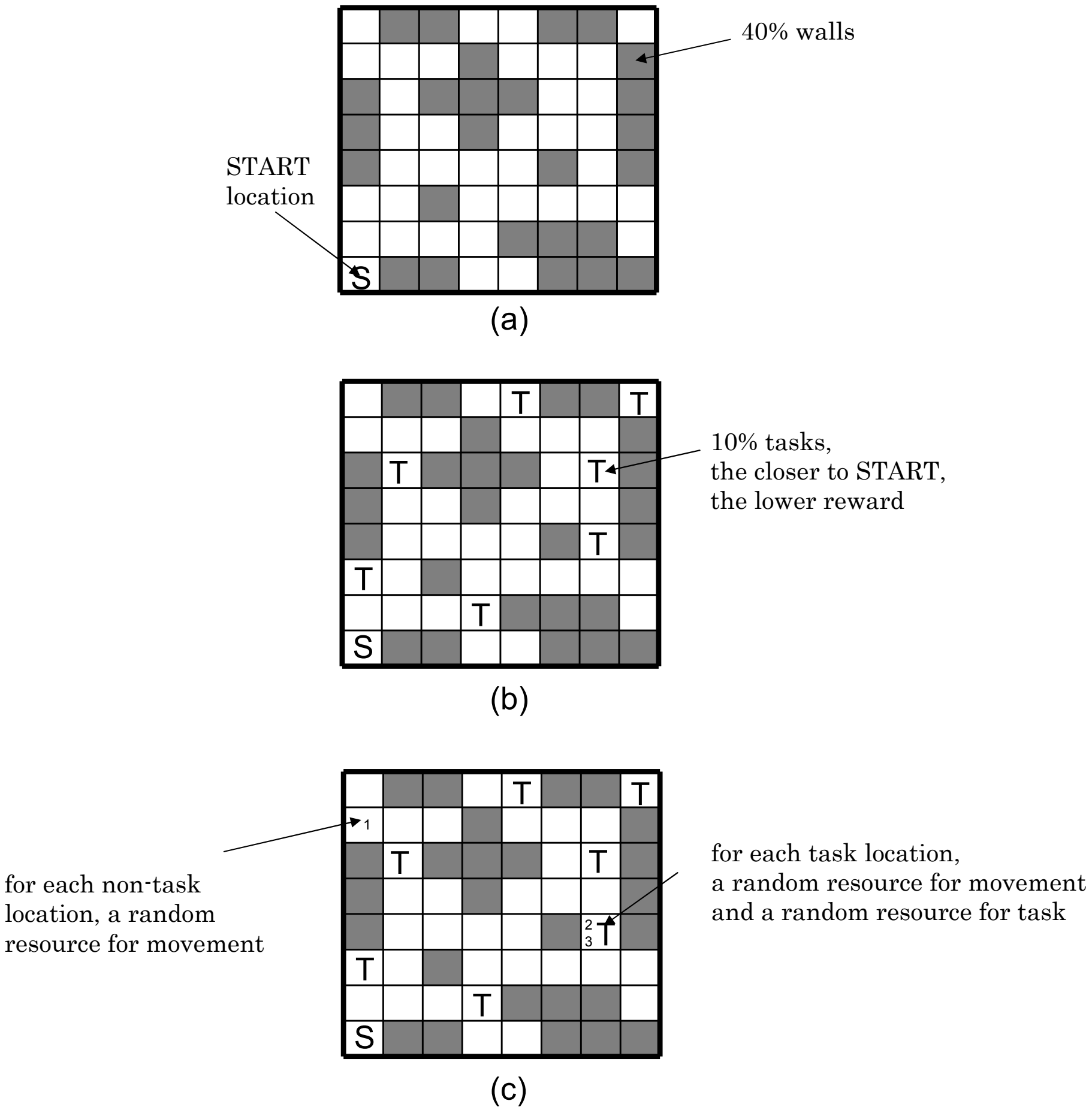

Figure 3: The procedure of creating a random grid world. (a) 40% of the locations are randomly chosen as walls. (b) 10% of the locations are randomly chosen for tasks. (c) resource requirements are randomly set.

The following lists the detailed action parameters used in our experiments:

***wait*** can be executed in any non-wall location without requiring any resource. After the execution of this action, the rover will stay at its current location with probability 0.95, and be out of the system with probability 0.05 (e.g., running out of battery).

***up*, *down*, *left*, *right*** can be executed in any non-wall location without requiring any resource. Each of these actions achieves its intended effect with probability 0.4, moves the rover into each of the other three directions (except the intended direction) with probability 0.1, keeps the rover in the current location with probability 0.1, and causes damage to the rover (and then the rover is out of the system) with probability 0.2. Furthermore, if the rover bumps into a wall, it will stay at its current location.

***safe-up*, *safe-down*, *safe-left*, *safe-right*** can be executed only in a location whose required resource is currently held by the rover. Compared to an unsafe-moving action, such a safe-moving action achieves the intended effect with a much higher probability 0.95, and fails (leaves the system) with a lower probability of 0.05. Similarly as before, when the rover bumps into a wall, it stays at its current location.

***do*** can be executed only for tasks whose required resources are currently held by the rover. When action *do* is executed, the rover receives the reward of the task at its current location, and leaves the system (since the mission is accomplished).

The capacity of the rover is restricted: the capacity limit is $\hat{\tau}$, and carrying each resource will incur one unit capacity cost. That is to say, the rover can carry no more than $\hat{\tau}$ resources.

We run experiments on a Core 2 Duo machine and use *CPLEX* 10.1 as our MILP solver. In our experiments, each average data point is computed from 20 randomly generated problems.

#### 3.5.2 Improvements to Solution Quality

We start the evaluation by showing the improved reward from using the phasing strategy over the approach that does not consider the possibility of switching resources in the midst of execution. Let us first consider the case where there are five supply stations distributed in the environment (the first station is always at the START location and the remaining four stations are randomly uniformly distributed in the non-wall locations when the problem is generated). Other parameters are set as follows: $n = 8$, i.e., the size of the grid world is 8 by 8, and $|O| = 9$, i.e., there are nine different types of resources in the system. Figure 4 gives average rewards for these experiments, where the error bars here and in the graphed results throughout this paper show the standard deviation. Clearly, exploiting the resource reconfiguration opportunities (using the abstract MDP solver presented in Section 3.3) can considerably improve the performance of the rover, e.g., receiving a reward on average about 40% higher than the reward when not taking advantage of the supply stations, when the rover can carry only three resources.

Figure 4 also compares the performance of the rover between the case where the locations of supply stations are randomly pre-selected and the case where the locations of the same number of supply stations (i.e., five phases, given that $\lambda_{i=START} = 0$, $\lambda_{i \neq START} = 1$, and $\hat{\lambda} = 4$) can be determined by the rover itself. As expected, finding optimal phase-switching states (which can be done by using the MILP algorithm presented in Section 3.4) is valuable in tightly constrained environments. For example, on average, it yields a reward about 46% higher than the approach that randomly selects phase-switching states when the number of carried resources is limited to $\hat{\tau} = 3$.

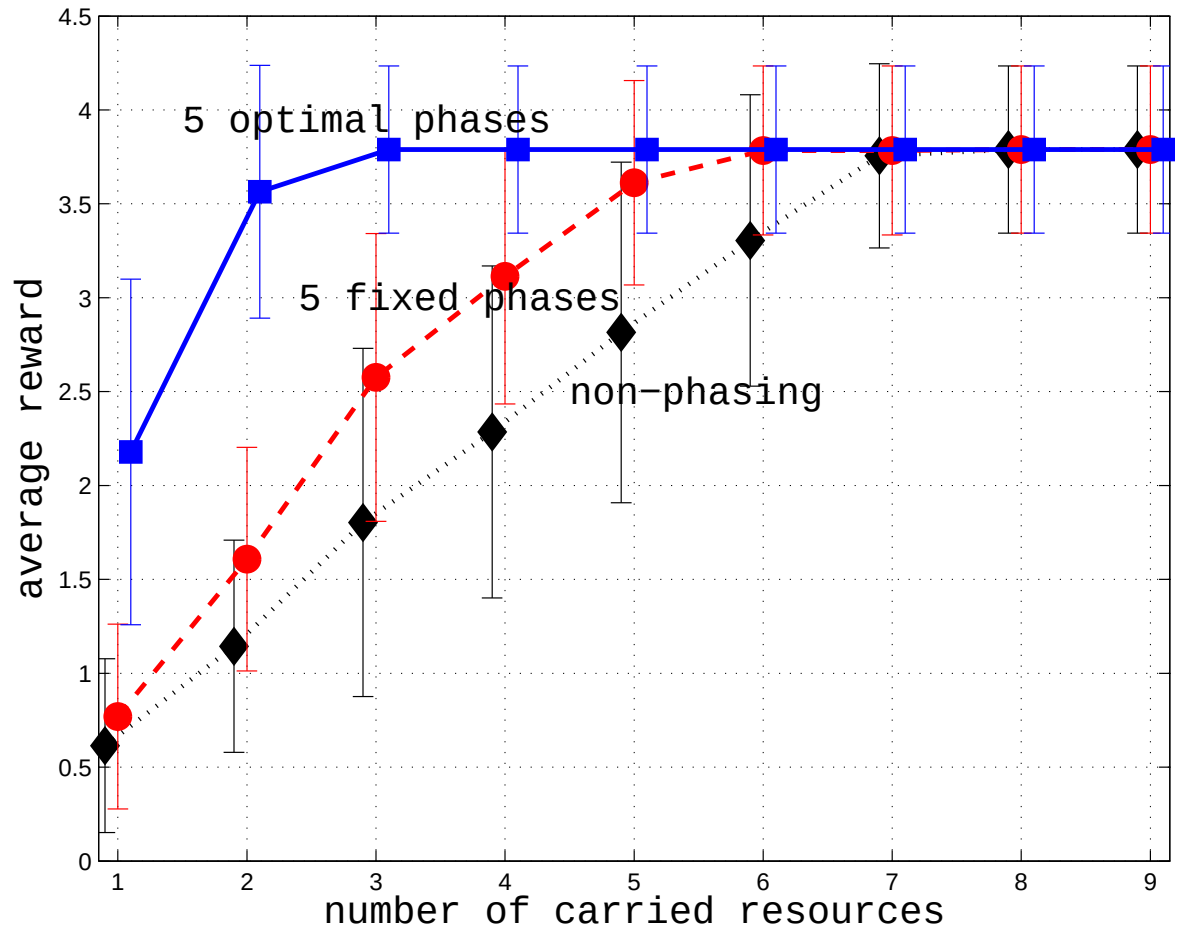

Figure 4: Exploiting fixed phase-switching states increases the agent's reward, and finding optimal phase-switching states further increases the reward.

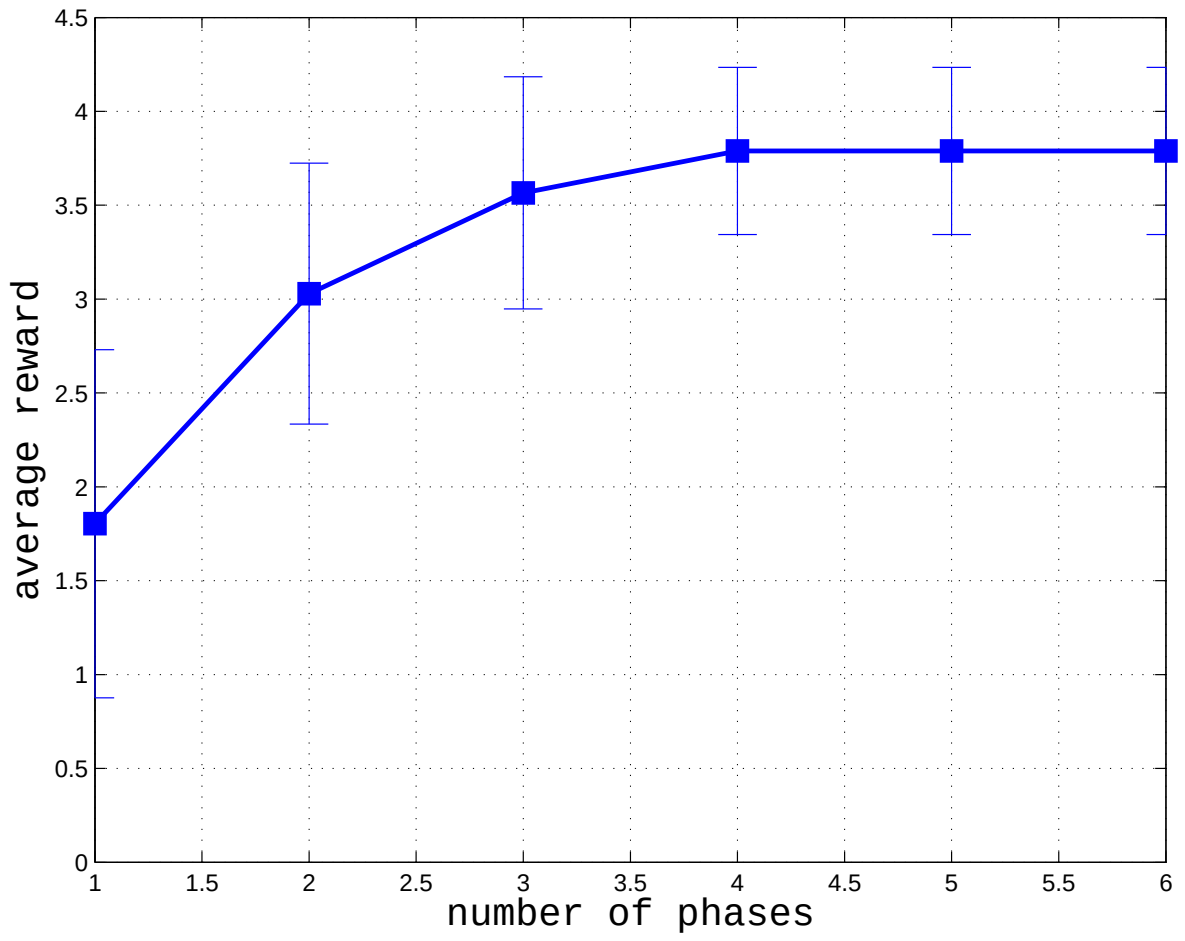

Figure 5: The reward increases as the number of phases increases.

Figure 5 examines the effectiveness of the resource-driven mission-phasing approach from another perspective, showing the average reward of the rover as a function of the number of phase-switching states that can be built in the environment (with other system parameters $n = 8$, $|O| = 9$, and $\hat{\tau} = 3$). We can see that (as expected) breaking the mission into multiple phases can significantly improve the total expected reward of the constrained rover. For example, setting up two additional supply stations in the $8 \times 8$ grid world environment (and so breaking it into three phases) can almost double the average reward that the rover can gain without using phasing.

#### 3.5.3 Computational Efficiency

A major objective of the work presented in this section is the design of a computationally-efficient solution approach for the S-RMP optimization problem. Section 3.2 has given a theoretical analysis of the computational complexity of the S-RMP optimization problem; this subsection is intended to empirically evaluate the efficiency of the solution approach presented in this section in solving complex S-RMP optimization problems. To make the presentation concise, only the runtime performance of the MILP-based algorithm described in Section 3.4 is shown, i.e., focusing on the standard S-RMP optimization problem defined in Section 3.1.[7]

The two prior straightforward algorithms described in Section 3.2 (the brute-force search approach based upon enumeration and the MDP-based approach incorporating resource features in the MDP state representation) can also be used to solve the S-RMP optimization problem. Enumerating all decompositions, and then, for each, enumerating all possible resource configurations and reconfigurations can be thought of as a (very slow) brute-force search algorithm for our formulated MILP. Therefore, we do not report its empirical results, since state-of-art MILP solvers (such as *CPLEX* which we use) usually follow more sophisticated branch-and-bound (B&B) strategies, and it is well established in the mathematical programming literature that the B&B approach is, in general, significantly better than the straightforward brute-force search (in both the runtime for finding an optimal solution and the anytime performance of finding a good solution). Our search of the Artificial Intelligence and the Operations Research literatures indicates that the MDP-based approach is the only existing approach (besides the brute-force search) that is directly applicable to the S-RMP optimization problem. We will thus focus on comparing our MILP-based algorithm and the MDP-based algorithm in the following discussion.

In the MDP-based algorithm, a new "drop-all" action and $|O|$ new "pick-one" actions are added into the original action space (instead of adding $2^{|O|}$ "resource reconfiguration" actions). That is to say, rather than performing resource reconfiguration in one step, the agent now switches to a new bundle of resources by first implementing a "drop-all" action and then sequentially performing "pick-one" actions until it has all its desired resources. Our experience is that this revised algorithm is more computationally efficient than the version with the exponential-size action space.

Note that the MDP resulting from incorporating resource features in the state representation is still a constrained MDP because phase-switching constraints place restrictions on which states resource-reconfiguration-related actions can be performed in. The constrained MDP solver (Eq. 4) has been shown to be efficient in solving large constrained MDPs (Dolgov, 2006), and so this work uses it for solving such remodeled constrained MDPs.[8]

7. Our experiments (Wu, 2008) also show that the trends of results for other variations of the S-RMP optimization problem are similar to those described in this subsection.
8. In the MILP formulation for solving the remodeled constrained MDP, the number of binary variables equals the number of states specified in the S-RMP problem definition. That is, the runtime of the MDP-based algorithm is exponential to the input size but not doubly exponential.

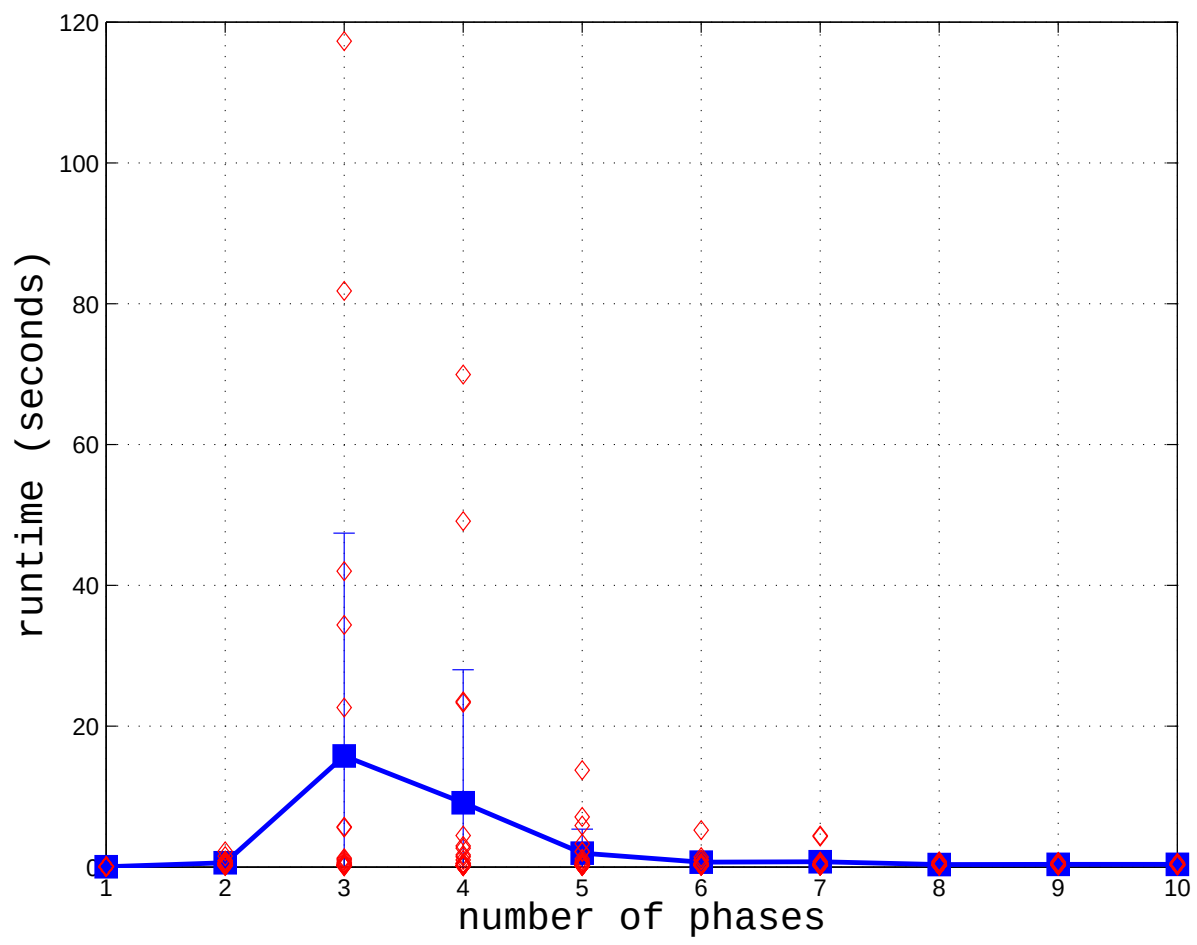

Figure 6: The runtime increases and then decreases as the number of phases increases.

To provide a better idea about the computational complexity of our experimental domain and solution techniques, we begin by showing what a "hard" resource-driven mission-phasing problem is, particularly along the dimension of the number of phases that can be created. We use the same parameters as in Figure 5, but analyze runtime instead. The results are shown in Figure 6, which demonstrates how the running time for deriving an optimal S-RMP solution varies as the number of supply stations that can be created in the environment increases. In the figure, the solid line shows the average, error bars show standard deviation, and each data point, which is shown as a "◇", corresponds to a single run.

As shown in the figure, the running time is low when the number of phases is small, and it gradually increases as the number of phases increases. This is not surprising, because the number of variables (both continuous variables and binary variables) in the MILP formulation is linear in the number of phases. However, the interesting observation is that, after some point, the runtime starts to decrease although the size of the MILP still keeps increasing. We believe this is because, when the number of allowable phases is large, there are several different ways to set up phase-switching states while achieving the same maximum reward. In other words, the S-RMP optimization problem with a large number of phase-switching states becomes under-constrained, and might have a large number of different optimal solutions. The MILP-based algorithm presented in this work can effectively exploit this property, and reduce computational costs. Based upon this complexity profile, to highlight the ability of solving "hard" problem instances, the following experiments set the phase-switching cost limit $\hat{\lambda}$ to 2 (except in the case where we examine how the running time changes as the number of phases increases). This means that there can be up to three phases in the system, assuming that creating each additional phase-switching state incurs one unit cost.

Figure 7 compares the average time for finding an optimal solution between our MILP-based algorithm and the standard MDP-based algorithm relative to the number of phases $\hat{\lambda}$ (top-left figure), the number of carried resources $\hat{\tau}_c$ (top-right figure), the number of resource types $|O|$ (bottom-left figure), and the size[9] of the grid world $n$ (bottom-right figure). We can

9. Recall the grid world is size $n \times n$.

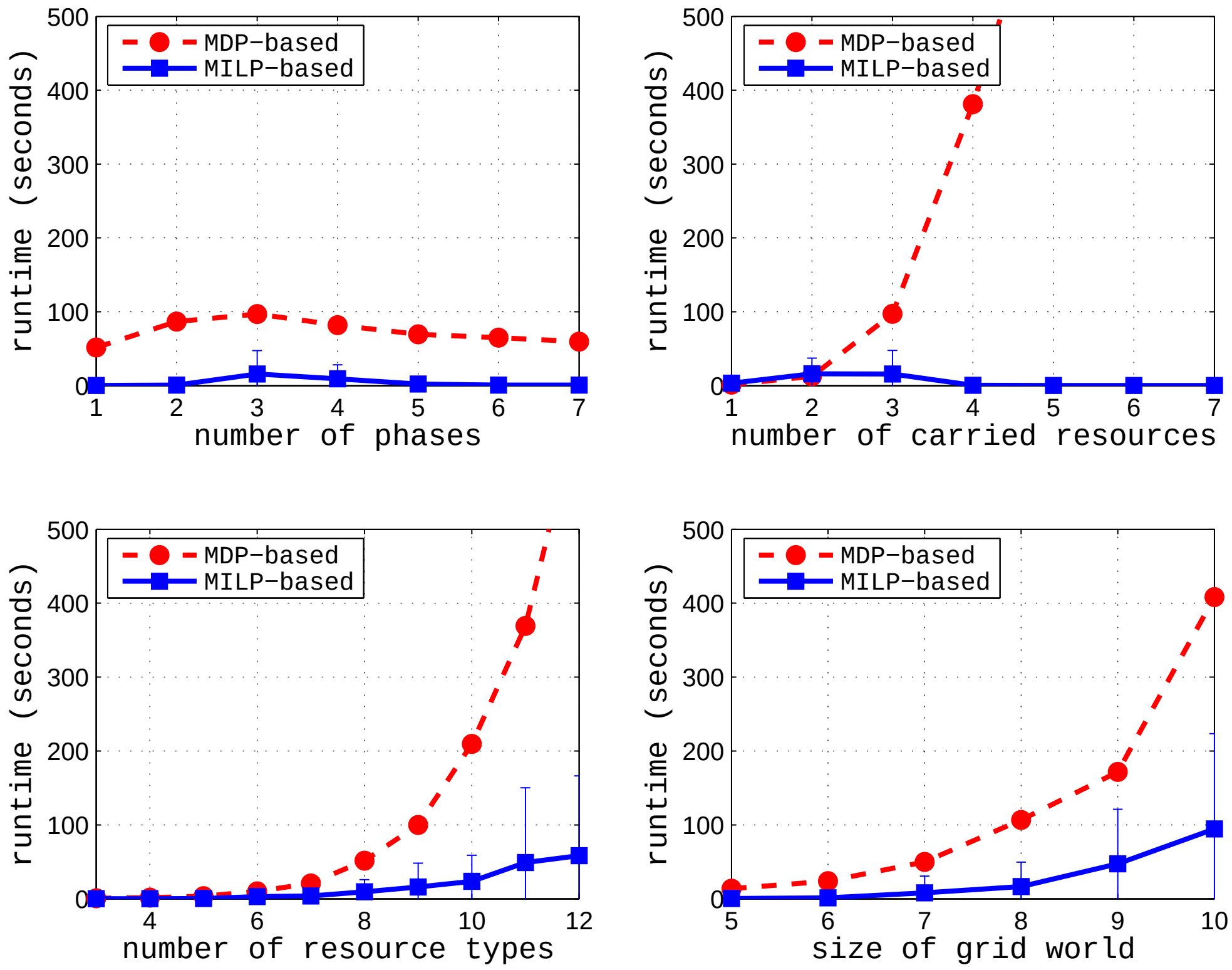

Figure 7: Runtime comparison between the MILP-based algorithm and the MDP-based algorithm. The MILP-based algorithm finds an exact solution to a S-RMP optimization problem faster than the standard MDP-based algorithm. Parameters are set as follows. Top-left figure: $n = 8$, $\hat{\tau} = 3$, $|O| = 9$, $\hat{\lambda} = \{0, 1.., 6\}$. Top-right figure: $n = 8$, $\hat{\tau} = \{1, ..., 7\}$, $|O| = 9$, $\hat{\lambda} = 2$. Bottom-left figure: $n = 8$, $\hat{\tau} = 3$, $|O| = \{3, 4, ..., 12\}$, $\hat{\lambda} = 2$. Bottom-right figure: $n = \{5, 6, ..., 10\}$, $\hat{\tau} = 3$, $|O| = 9$, $\hat{\lambda} = 2$.

see that our MILP-based algorithm is in expectation considerably faster than the MDP-based algorithm, particularly in complex problem instances.

In the top-left figure, the results of the MILP-based algorithm are the same as those shown in Figure 6, which have already been discussed. Notably, unlike the other three figures, the curve of the MDP-based algorithm in this figure does not monotonically increase as the value of the input parameter increases. This is because the input parameter in this figure, the number of phases, does not affect the size of the state space of the expanded MDP. Furthermore, the constrained MDP method (Eq. 4) used to solve the expanded MDP can exploit problem structure when the problem becomes under-constrained. This explains why the running time decreases after some point (but the time is still much higher than that of the MILP-based approach).

The top-right figure also demonstrates a trend for the running time of the MILP-based algorithm decreasing after the value of the input parameter (i.e., the number of resources that can be carried by the rover) is above a particular threshold. The reason is the same as that used to explain Figure 6 — the MILP-based algorithm, where the MILP solver utilizes branch-and-bound, can effectively discover and exploit the fact that the problem becomes under-constrained. In contrast, the MDP-based algorithm incorporating resource features into the state representation leads to a MDP whose size grows very rapidly as the number of resources that can be carried increases, and thus results in a significant increase in the running time.

As illustrated in the bottom-left figure and the bottom-right figure, the average runtime of the MILP-based algorithm also increases considerably more slowly than the MDP-based algorithm, although, unlike the top-left and top-right figures, the runtime monotonically increases as either the number of resource types or the size of the grid world increases. This is because, in general, the increases of these two parameters will not make the problem become under-constrained by themselves.

The reason for the significant reduction in computational cost is that our MILP-based approach can formulate the S-RMP optimization problem in a compact (as opposed to exponential) formulation, which paves the way for taking advantage of state-of-the-art MILP solvers to effectively solve the coupled problems of problem decomposition, resource configuration, and policy formulation. It is important to emphasize that the MILP-based approach uses no approximation techniques (and so it will find optimal solutions). The compactness of the formulation is because the MILP-based approach folds the process of solving a NP-complete S-RMP problem into the process of solving a NP-complete MILP (where the MILP can be solved efficiently by state-of-the-art solvers).

Specifically, the MDP-based approach models resources in the MDP representation regardless of valuations of subsets of the resources, and then it reasons over the generalized MDP to determine an optimal way of configuring and reconfiguring resources. In contrast, our MILP-based solver finds an exact S-RMP solution by taking advantage of the embedded branch-and-bound MILP method to discard subsets of fruitless candidate solutions (through upper and lower estimated bounds). Although the MILP-based approach and the MDP-based approach have similar worst-case runtimes, i.e., requiring exponential time to enumerate all possible ways of sequentially configuring resources (which is reasonable because S-RMP is NP-complete), the average-case performance of the MILP-based approach is often much better than the MDP-based approach because of the effectiveness of the branch-and-bound algorithm for pruning suboptimal solutions. This is particularly significant in cases where suboptimal decompositions can be detected easily and early because a large number of possible resource configurations and executable policies can then be discarded without much computational effort.

### 3.6 Summary

To this point, we have analyzed several variations of a single-agent resource-driven mission-phasing problem, corresponding to several cases of phase-switching constraints, and presented a suite of computationally efficient algorithms for finding and using mission phases. We have shown through analysis and experiments that our approach can considerably reduce the computational cost for finding an exact solution to a complex S-RMP optimization problem in comparison with prior approaches. In the remainder of this paper, we will extend such techniques into multiagent stochastic systems.

## 4. Resource Reallocation in Multiagent Systems

Additional complications arise when an agent that is deciding which resources to hold is part of a multiagent system, because of potential competition for scarce, shared resources. For example, there might be only be a few satellites to remotely control to acquire desired images, or a small number of licenses to simultaneously run a software package. An individual agent might be unable to procure all of its desired resources (even when its capacity does not restrict the amount of resources it can hold) because some other agents may want those resources as well. For cooperative agents, an optimal allocation will distribute resources to agents so as to maximize the agents' aggregate reward, meaning that a shared resource (such as control of a satellite's sensor) should be given to the agent that can use it best, given its goals, potential actions, and other resources possessed.

This problem is in general doubly-exponential. Not only could each agent need to consider exponentially many combinations of resources that it might possess, but then collectively the agents could need to consider the exponentially many joint combinations of their individual combinations, filtering out those that exceed shared resource constraints, and returning the best of those that remain. Dolgov and Durfee (2005, 2006) showed that a much more efficient algorithm is possible that exploits problem structure for the case where an agent's valuation for a resource bundle is based on the expected value for an MDP policy that utilizes the actions made possible by holding the resources in the bundle. We will summarize that work in Section 4.2, but the approach is similar to their solution method for resource assignment to a single agent with capacity constraints described in Section 2.3, and like that work solves the *one-shot* allocation problem. Hence, like that work it does not consider the possibility that the agents might be able to redistribute resources among themselves in the midst of mission execution.

This section thus focuses on solving *sequential* resource redistribution (reallocation) problems, along with the problem of optimizing agents' policies for the execution phases between redistribution events, by building on ideas from Dolgov and Durfee's work as well as the S-RMP techniques presented in the previous section. The remainder of this section thus largely follows a parallel structure to the preceding S-RMP presentation. We begin (Section 4.1) by introducing a simple problem that we will use for running examples through the remainder of the section. Then we will summarize (and using the example illustrate) the prior work of Dolgov and Durfee (2005, 2006) on one-shot allocation in Section 4.2, and define the sequential multiagent resource-driven mission phasing problem in Section 4.3. After analyzing the problem's complexity (Section 4.4), we then consider a sequence of variations on the problem (again paralleling the S-RMP description), beginning with the case where the phase-switching states are pre-defined (Section 4.5) and then where they can be chosen (Section 4.6), and for each, present, analyze, and illustrate solution algorithms. Both efficiency and optimality of our techniques are empirically evaluated in Section 4.7. Finally, Section 4.8 summarizes the contributions of the work presented in this section.

### 4.1 A Multiagent Example

We here describe a simple multiagent resource-allocation example problem that we will use to illustrate the various solution approaches throughout this section. In this example, two cooperative agents attempt to maximize their total expected reward over a ten time step interval. Each agent has three tasks. At each time step, an agent can choose to continue its previously started task (if there is one and if the required resources are still assigned to that agent), to start a new task (and abort its current task if there is one), or simply to do nothing. In addition, we say that a task that has been aborted previously (and thus has failed) can be re-tried, but no task can be accomplished more than once.

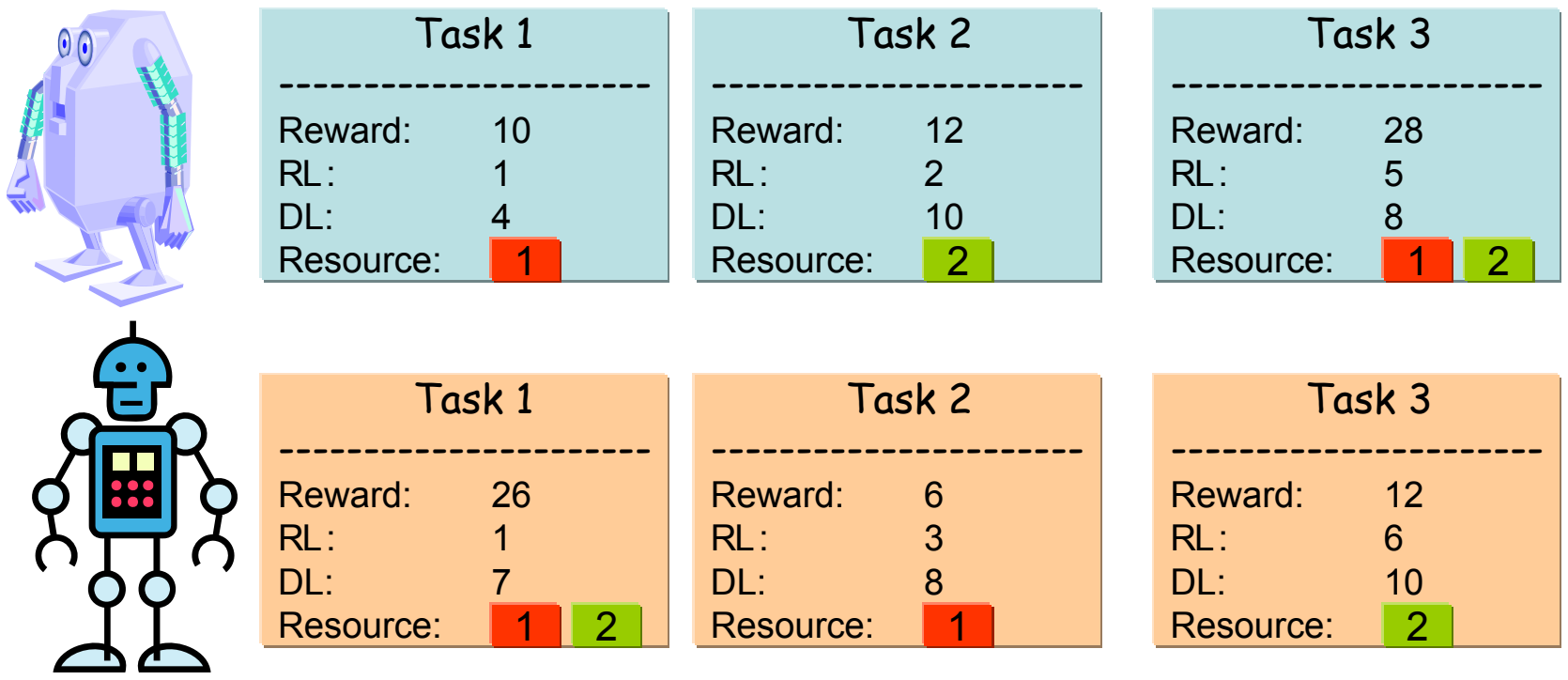

Figure 8: A simple two-agent example.

Figure 8 shows the detailed information of the tasks in the example problem, including release (RL) time (i.e., the earliest time the task can be started successfully), deadline (DL) (i.e., the latest time the task can finish successfully), reward if the task completes successfully, and resource prerequisites. For example, agent *1* can start (or continue) its task *1*, which will incur a reward 10 if accomplished, at any time step within the interval [1, 4) given that it has one unit of resource *1* at that time. To concentrate on the multiagent issues, for this problem we will assume that each agent has sufficient capacity to carry both resources *1* and *2*. The uncertainty in this problem is in the amount of time required to execute a task. Here, we say that, if an agent starts a task and does not abort it during execution, then the agent has probability 0.3, 0.4, and 0.3 of accomplishing it within one, two, and three time steps, respectively.[10]

When there are multiple instances each of resources *1* and *2*, then this problem degenerates to two independent unconstrained MDP problems, one for each agent. As illustrated in Figure 9, each agent will hold both resources it needs throughout the full time interval. Using a standard policy formulation algorithm (e.g., value iteration), we can easily compute the optimal unconstrained policy for each agent, which yields a total expected reward across the two agents of 93.64.

### 4.2 Background: Integrated Resource Allocation and Policy Formulation

Stochastic planning in multiagent environments is typically much more challenging than in single-agent environments, particularly when each agent has only a partial view of the global environment. Previous complexity analyses have shown that the general decentralized Markov decision process (Dec-MDP) is NEXP complete (Bernstein, Zilberstein, & Immerman, 2000;

10. This problem representation is a simplification of the TAEMS modeling approach (Lesser, Decker, Wagner, Carver, Garvey, Horling, Neiman, Podorozhny, Prasad, Raja, Vincent, Xuan, & Zhang, 2004; Wagner, Raja, & Lesser, 2006) as used in the DARPA COORDINATORS project (Wagner, Phelps, Guralnik, & VanRiper, 2004; Wu & Durfee, 2007b).

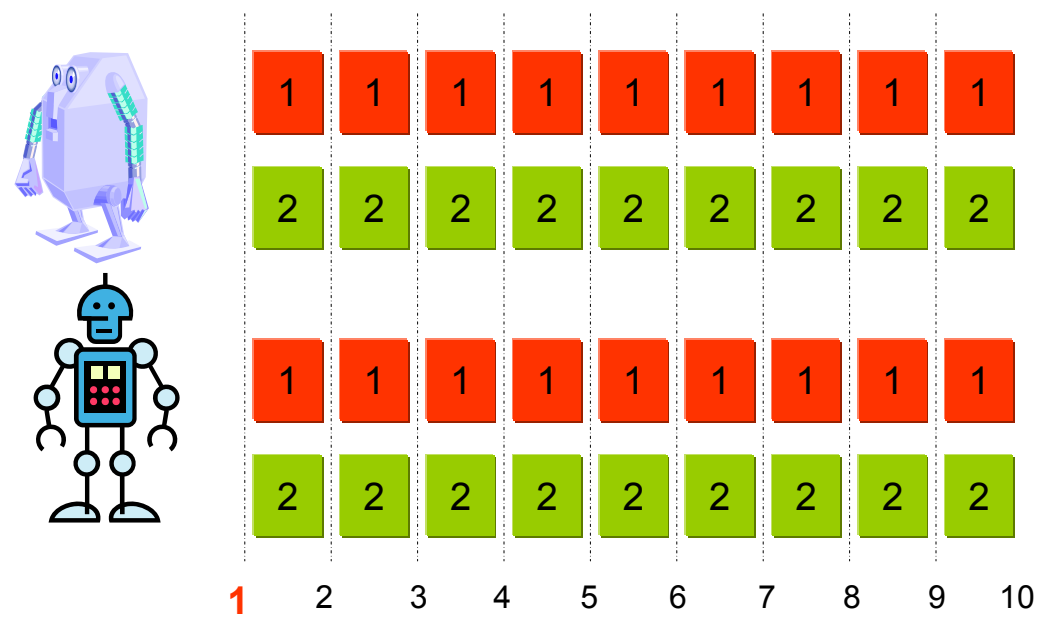

Figure 9: Optimal resource allocation when resources are unlimited.

Goldman & Zilberstein, 2004). Fortunately, in many application domains, the actions taken by one agent may not impact other agents' transitions. For example, when a few delivery robots operate in a large open environment, interactions may be rare and easily avoidable. The development of efficient algorithms for such *loosely-coupled* systems has gained much attention among many researchers (e.g., Meuleau, Hauskrecht, Kim, Peshkin, Kaelbling, Dean, & Boutilier, 1998; Becker, Zilberstein, Lesser, & Goldman, 2004; Dolgov & Durfee, 2005).

Our approach specifically draws upon the prior approach designed by Dolgov and Durfee (2005, 2006), which extends their work summarized in Section 2.3 to the multiagent resource allocation case. Their work assumes that a group of cooperative agents are coupled through sharing resources (i.e., actions selected by one agent might restrict the actions available to others), but that the actions executed by one agent cannot impact the rewards and transitions of others. As is typical in the resource-allocation research literature, the work also assumes that, once the resources are distributed, the utility that each agent can achieve is only a function of its own resource assignment and does not depend on what resources are given to other agents and how they use their resources.

A multiagent constrained MDP with scarce shared resources can be represented as the tuple $\langle \mathcal{M}, \alpha, \mathcal{C} \rangle$, as next specified. Note that the specification essentially represents an independent instance of a constrained MDP (Section 2.3) for each agent, except that $O$ (the set of resources) and $\hat{\Omega}$ (the bounds on the number copies of each resource) are shared across all agents.

- $\mathcal{M} = \{\mathcal{M}^m\}$ is a set of classical MDPs, where $\mathcal{M}^m$ represents agent $m$'s MDP, modeled in the same way as described in Section 2.1. That is, $\mathcal{M}^m = \langle S^m, A^m, \{p^m_{i,a,j}\}, \{r^m_{i,a}\} \rangle$ where $S^m$ is the finite state space of agent $m$, $A^m$ is the finite action space of agent $m$, $p^m_{i,a,j}$ is the probability that agent $m$ reaches its state $j$ when it executes action $a$ in its state $i$, and $r^m_{i,a}$ is the reward agent $m$ receives when it performs action $a$ in its state $i$.

- $\alpha = \{\alpha^m\}$ specifies the initial probability distribution over states for each agent $m$, where $\alpha^m_i$ is the probability that agent $m$ is initially in its state $i$.

- $\mathcal{C}$ represents resource constraints on agents in the system, which can be represented as $\langle O, C, U, \Gamma, \hat{\Gamma}, \hat{\Omega} \rangle$ where:

◇ $O$ is the finite set of shared, indivisible, non-consumable execution resources.[11]

◇ $C = \{C^m\}$ specifies the finite set of capacities for each agent $m$.

◇ $U = \{U^m\}$ gives the resource requirements for each agents' actions, where the binary parameter $u^m_{o,a,i} \in \{0, 1\}$ indicates whether agent $m$ requires resource $o$ to execute action $a$ when it is in its state $i$.

◇ $\Gamma = \{\Gamma^m\}$ represents the capacity costs for each agent $m$.

◇ $\hat{\Gamma} = \{\hat{\Gamma^m}\}$ captures the capacity limits for each agent $m$.

◇ $\hat{\Omega} = \{\hat{\omega}_o\}$ specifies (shared) resource limitations, where $\hat{\omega}_o$ is the maximum number of copies of resource $o$ that can be distributed to agents in the system.

**Running Example: Multiagent Constraint Formulation.** *In the simple running example from Section 4.1, the agents' constraint components $\mathcal{C}$ are summarized below, where $*$ represents any state and unspecified values of $u$ are zero.*

- $O = \{o_1, o_2\}$.
- $C = \{\{hold\}, \{hold\}\}$.
- $U = \{\{u^1_{o_1,a_1,*} = 1, u^1_{o_2,a_2,*} = 1, u^1_{o_1,a_3,*} = 1, u^1_{o_2,a_3,*} = 1\}, \{u^2_{o_1,a_1,*} = 1, u^2_{o_2,a_1,*} = 1, u^2_{o_1,a_2,*} = 1, u^2_{o_2,a_3,*} = 1\}\}$.
- $\Gamma = \{\{\tau^1_{o_1,c_{hold}} = 1, \tau^1_{o_2,c_{hold}} = 1\}, \{\tau^2_{o_1,c_{hold}} = 1, \tau^2_{o_2,c_{hold}} = 1\}\}$.
- $\hat{\Gamma} = \{\{\hat{\tau}^1_{c_{hold}} = 2\}, \{\hat{\tau}^2_{c_{hold}} = 2\}\}$.
- $\hat{\Omega} = \{\omega_{o_1} = 1, \omega_{o_2} = 1\}$.

The algorithm devised by Dolgov and Durfee (2006) is presented below in Eq. 8, for clarity leaving out the capacity constraints to focus on the multiagent constraints. Analogously to the single-agent case, the continuous variable $x^m_{i,a}$ represents the expected number of times agent $m$ executes action $a$ in its state $i$, and the binary variable $\Delta^m_o$ represents whether one unit of resource $o$ is assigned to agent $m$ prior to execution.

$$\max \sum_m \sum_i \sum_a x^m_{i,a} \times r^m_{i,a} \tag{8}$$

11. For simplicity, we will here assume that all resources are shared. If agents can also draw from a cache of private resources then this is a straightforward extension but would add unnecessary complication to the presentation.

subject to:

*probability conservation constraints:*

$$\sum_a x^m_{j,a} = \alpha^m_j + \sum_i \sum_a p^m_{i,a,j} \times x^m_{i,a} \qquad : \forall m, \forall j$$

$$x^m_{i,a} \geq 0 \qquad : \forall m, \forall i, \forall a$$

*resource constraints:*

$$\frac{\sum_i \sum_a u^m_{o,a,i} \times x^m_{i,a}}{X} \leq \Delta^m_o \qquad : \forall m, \forall o$$

$$\sum_o \tau^m_{o,c} \times \Delta^m_o \leq \hat{\tau}^m_c \qquad : \forall c, \forall m$$

$$\sum_m \Delta^m_o = \hat{\omega}_o \qquad : \forall o$$

$$\Delta^m_o \in \{0, 1\} \qquad : \forall m, \forall o$$

- The objective function $\sum_m \sum_i \sum_a x^m_{i,a} \times r^m_{i,a}$ represents the sum of cumulative rewards among all agents, based upon the assumption that the agents are coupled only through resources (i.e., actions taken by one agent will not impact other agents' rewards and transitions).
- The constraint $\sum_a x^m_{j,a} = \alpha^m_j + \sum_i \sum_a p^m_{i,a,j} \times x^m_{i,a}$ guarantees probability conservation at every state for every agent, which is a multiagent version of the probability conservation constraint in the single-agent MDP formulation (Eq. 1).
- $X$ is a constant equal to or greater than $\sup \sum_i \sum_a x^m_{i,a}$ (where in a finite horizon MDP, $X$ can be set to the finite horizon $T$ since each agent can only execute $T$ actions within that horizon). The constraint $\frac{\sum_i \sum_a u^m_{o,a,i} \times x^m_{i,a}}{X} \leq \Delta^m_o$ implies that $x^m_{i,a}$ must be zero (i.e., action $a$ cannot be executed by agent $m$ in state $i$) when $u^m_{o,a,i} = 1$ (i.e., agent $m$ must have resource $o$ to execute action $a$ in its state $i$) and $\Delta^m_o = 0$. $x^m_{i,a}$ is unrestricted otherwise since $X$ is no less than $\sum_i \sum_a u^m_{o,a,i} \times x^m_{i,a}$ by definition.
- The constraint $\sum_m \Delta^m_o = \hat{\omega}_o$ guarantees that the total amount of resource $o$ allocated across all agents must equal the amount of available resource $o$ (assuming the resources will be completely assigned). This constraint can be easily relaxed to the constraint $\sum_m \Delta^m_o \leq \hat{\omega}_o$ by introducing an additional *dummy* agent to keep unallocated resources.

The optimal joint policy can be easily derived from the solution to the above MILP in a similar way to that discussed in Section 2.2. That is, to maximize the total expected reward of the group of agents, agent $m$ should choose $a$ with probability $\frac{x^m_{i,a}}{\sum_a x^m_{i,a}}$ when it is in its state $i$.

**Running Example: Multiagent Constrained MDP Solution.** *Continuing the simple running example from Section 4.1, we can apply the above solution technique to find an optimal allocation of resources* 1 *and* 2 *at the outset to the agents assuming that there is only one copy of each available. Eq. 8 finds that the optimal one-shot allocation is to give all resources to agent* 1 *and let agent* 2 *idle over the entire execution, as shown in Figure 10, and the total*

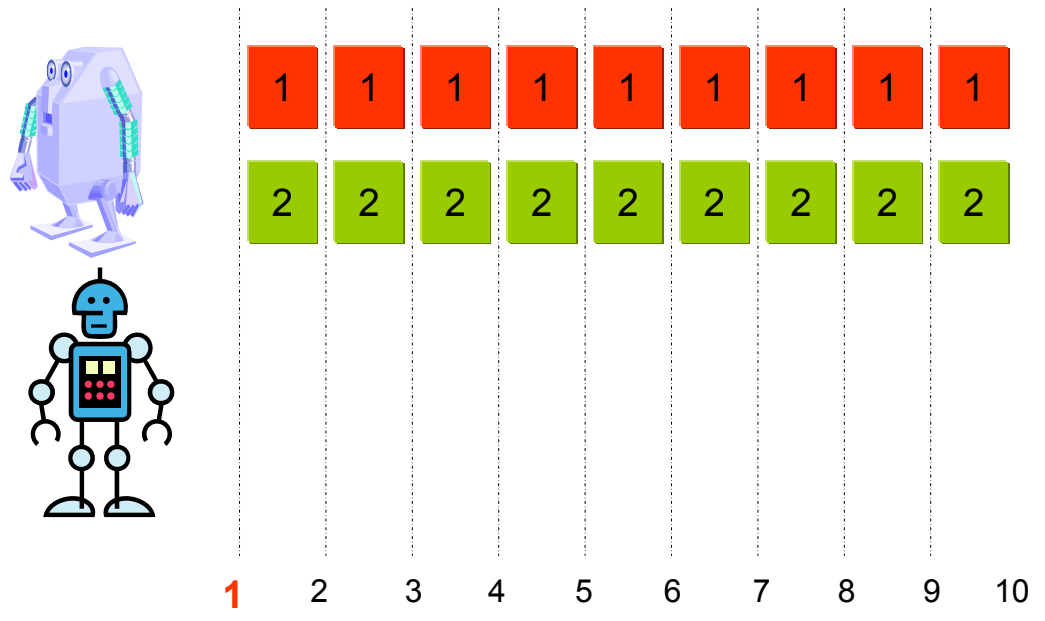

Figure 10: Optimal one-shot resource allocation when resources are scarce.

*expected reward is 49.64 in this case, much lower than the reward (93.64) in the unconstrained case.*

### 4.3 Multiagent Resource-Driven Mission Phasing Problem Definition

The Multiagent Resource-driven Mission Phasing (M-RMP) problem is the sequential version of Dolgov and Durfee's (one-shot) multiagent resource allocation problem as just described. A general multiagent mission-phasing problem can be solved exactly by using the S-RMP solution approach presented in Section 3 on the joint state and action spaces of the interacting agents (assuming that each agent has a full view of the joint state), but such a solution methodology would suffer from the curse of dimensionality since the sizes of the joint state and action spaces grow exponentially with the number of agents. Thus, we will exploit the loose-coupling assumption: agents only interact through their (now potentially repeated) contention for shared resources, and are otherwise transition and reward independent.

In the multiagent case of sequential resource (re)allocation, each phase corresponds to a particular distribution of shared resources among the agents. We will say that the transition from one phase into another occurs when all of the agents reach a state where they relinquish possession of the resources they currently hold, and then acquire their resources for the next phase.[12] However, unlike the single-agent case where the agent knows when it has reached a phase-switching state, in the multiagent case each agent can only observe its local state, and so detecting a joint phase-switching state would require agents to communicate state information and individually track the full joint state. This in turn means that each would need to have a policy that maps the (exponentially larger space of) joint states to local actions, effectively introducing transition and reward dependencies and exploding the complexity of the problem.

Our work instead has the agents exploit a deterministically-changing feature of the joint state that they all can observe: a (synchronized) clock. Phase switching—redistributing resources such as software licenses or control of a satellite's sensors—occurs at pre-arranged (and preferably carefully chosen) times. We already saw in Section 3.4.2 a (single-agent) RMP

12. While in principle different subsets of agents could swap resources among themselves, in this paper we only consider the case where *all* agents engage in swaps, though of course since an agent can acquire the same resources it relinquishes the effect could be that only a subset of the agents is materially involved in any given swap.

variation where phase-switching states were grouped based on the value(s) of a (subset of) feature(s). M-RMP builds on that variation, effectively partitioning the exponentially-sized joint state space into subsets based on the states' time feature, and determining the time(s) at which resource (re)distributions will take place, no matter what the other details of the agents' states are or whether they are all finished using their currently-held resources yet.

Phasing based only on time has both advantages and limitations. One key advantage is that all it requires that agents know at runtime about each other is that they all know what time it is (and that this is common knowledge). In domains where reliable mutual observation or communication at runtime is impractical (for example, in some military operations), synchronizing actions based on clock time has long been the norm. A second advantage is that a future time is guaranteed to be reached. In contrast, if our agents need other conditions to be met to exchange resources (for example, they need to be in the same location and/or all be idle), then in some applications it might be impossible to guarantee that such a state will ever occur. Or to force such cases to occur, agents would need to know more about each others' states (where others are and/or how soon their current tasks will end) to reach a hand-off point. The M-RMP techniques we describe can in principle be extended to such cases, but in practice the greater the need for agents to have global state awareness, the lesser we expect such problems to exhibit the kinds of structure that M-RMP exploits.

For this paper, we also assume that resource redistributions always succeed (resources do not somehow get "misplaced" during a transfer); our discussion of future work (Section 6.2) will talk about the implications of relaxing this assumption.

Based on the preceding, we now formally define the *multiagent resource-driven mission-phasing* (M-RMP) problem. It is a generalization of the multiagent constrained MDP described in Section 4.2, and is a tuple $\langle \mathcal{M}, \alpha, \mathcal{C}, \text{ and } \mathcal{R} \rangle$, where the components $\mathcal{M}$, $\alpha$, and $\mathcal{C}$ are as defined in Section 4.2, and $\mathcal{R}$ is defined as follows.

- ☐ $\mathcal{R}$ specifies constraints on resource reallocation. We capture the efforts required for such resource reallocation activities as costs $\langle \{\psi_t\}, \hat{\psi} \rangle$:[13]
  - ◇ $\{\psi_t\}$ indicates resource reallocation costs, where $\psi_t$ denotes the cost for reconfiguring the resource assignment at time $t$. Note that $\psi_t$ is only associated with time $t$ regardless of what resources and how many of them are reassigned. A variation of resource reallocation constraints where the reallocation cost depends on the amount of resources being transferred will be discussed and analyzed in Section 4.6.1 after presenting the solution algorithm to the M-RMP optimization problem defined in this section.
  - ◇ $\hat{\psi}$ specifies the limit on the amount of cost that could be spent in resource reallocation. For example, $\psi_{t=any_time} = 1$ and $\hat{\psi} = 4$ means that at most four resource reconfiguration events could be scheduled during a particular mission execution.

**Running Example: M-RMP Formulation.** *Continuing the simple running example from Section 4.1 and building on the encoding from Section 4.2, the agents' reallocation constraints $\mathcal{R}$ are summarized below, assuming the agents can reallocate resources three times besides at the initial time 1.*

13. This is different from *buffer pool* research (Lehman & Carey, 1986; Sacco & Schkolnick, 1982), which often assumes that buffer size can be changed immediately and free of charge.

- $\psi = \{\psi_1 = 0,\ \psi_i = 1 : 2 \leq i \leq 10\}$.
- $\hat{\psi} = 3$.

The objective of the M-RMP optimization problem is to maximize the total expected reward of a group of agents within a finite time horizon by judiciously reallocating the limited, shared resources among the agents over time. Although much simpler than a general decentralized MDP problem, such an automated multiagent mission-phasing problem is still computationally challenging because it needs to determine not only how to initially allocate limited shared resources, but also when to reallocate resources, what the best way of reallocating resources at those times are, and what the best executable policies with respect to the reallocated resources are. As in S-RMP, these three component problems — mission decomposition, resource allocation, and policy formulation — are strongly intertwined. The utility of decomposing a problem into phases and the utility of allocating resources for each phase are unknown until executable policies are formulated and evaluated, but the policies cannot be formulated until the phases are built and the resources are allocated.

### 4.4 Computational Complexity Analysis

This section starts by theoretically analyzing the computational complexity of the M-RMP optimization problem.

**Theorem 4.1.** *M-RMP optimization is NP-complete.*

*Proof:* It is trivial to prove that the M-RMP optimization problem is NP-hard. Given that its special case — one-shot resource allocation and policy formulation — can be proven to be NP-complete through a reduction from the KNAPSACK problem (Dolgov, 2006), M-RMP optimization is NP-hard.

Given a solution to the M-RMP problem, the satisfaction of resource constraints and resource reallocation constraints can be verified in linear time. After that, for each agent, incorporating its policy into its MDP model, the M-RMP optimization problem becomes a Markov chain, which can be solved in polynomial time. That is, M-RMP optimization is in NP.

With both NP and NP-hard, M-RMP optimization is NP-complete. □

Given this result, it is not surprising that the prior approaches (summarized below) that could be applied to finding an exact solution to the M-RMP optimization problem are not computationally efficient.

**Decentralized MDP.** Modeling resources into the MDP state representation and formulating resource-reconfiguration activities as actions is a possible (but slow) way to solve a S-RMP problem (Section 3), but is generally infeasible for the M-RMP problem. Since the outcome of an agent's resource-reallocation action (e.g., acquiring a resource) depends on whether another agent takes a corresponding action (e.g., releasing a resource) before or at the same time, the resulting Decentralized MDP (Dec-MDP) is not transition independent. A general Dec-MDP is NEXP-complete (Bernstein et al., 2000), meaning the M-RMP involves solving a NEXP-complete problem with an input exponential in the number of resources.

**Combinatorial and stochastic optimization.** Although each phase in a M-RMP problem is a one-shot resource-allocation and policy-formulation problem, directly using the integrated combinatorial and stochastic optimization approach (Section 4.2) to solve each phase independently and then piecing these phase policies together is, in general, infeasible. Besides having to enumerate all possible decompositions, solving each phase independently requires knowing the initial state probability distribution $\alpha_j^m$ for Eq. 8. Unfortunately, $\alpha_j^m$ of a phase generally depends on the policy of its preceding phase, but the policy of a preceding phase usually can only be optimized with respect to already knowing the expected utilities attainable in the current and future phases.

**Auction-based resource allocation.** In a resource allocation approach based on using auctions (Pekec & Rothkopf, 2003; De Vries & Vohra, 2003), each agent bids a set of valuations over its possible sequential resource assignments to a central auction, which then decides how to sequentially allocate resources among the agents. Unfortunately, this approach does not scale. For example, if a group of $m = 5$ agents wants to maximize the total expected reward within $T = 10$ time steps, (re)distributing $o = 5$ shared resources at most $k = 3$ times (twice after the initial allocation), then each agent needs to solve $C_{k-1}^{t-1} \times (2)^{o \times k} = 1,179,648$ non-trivial problems to evaluate all possible sequential resource assignments. Then, the auction faces a winner determination problem (WDP) where each of the 5 agents submits $1,179,648$ bids – a daunting task.

As in the S-RMP problem, our solution is to formulate the problem so as to simultaneously solve the coupled problems of mission decomposition, resource allocation, and policy formulation, to exploit interactions among them and to reduce computational cost.

### 4.5 Exploiting a Fixed Resource Reallocation Schedule

As in Section 3.3, we begin with a simple variant of the problem, where the schedule of reallocating resources is predetermined, i.e., resource reallocation cost $\psi_t = 0$ if time step $t$ is specified in a predefined schedule, $\psi_t > 0$ otherwise, and the cost limit $\hat{\psi} = 0$.

Section 4.4 explained why directly applying the (one-shot) integrated combinatorial and stochastic optimization approach to each phase independently and then piecing phase policies together is generally infeasible. Our approach instead links the phases together by modeling transition probability conservation. The details are given in the following MILP. Note that, to highlight M-RMP's emphasis on resource (re)allocation, we continue here and throughout the remainder of this section to omit the capacity constraints.

$$\max \sum_m \sum_i \sum_a x_{i,a}^m \times r_{i,a}^m \tag{9}$$

subject to:

*probability conservation constraints:*

$$\sum_a x^m_{j,a} = \alpha^m_j + \sum_i \sum_a p^m_{i,a,j} \times x^m_{i,a} \qquad : \forall m, \forall j$$

$$x^m_{i,a} \geq 0 \qquad : \forall m, \forall i, \forall a$$

*resource constraints:*

$$\frac{\sum_{i \in \mathbb{S}^k} \sum_a u^m_{o,a,i} \times x^m_{i,a}}{T} \leq \Delta^{m,k}_o \qquad : \forall k, \forall m, \forall o$$

$$\sum_m \Delta^{m,k}_o = \hat{\omega}_o \qquad : \forall o, \forall k$$

$$\Delta^{m,k}_o \in \{0, 1\} \qquad : \forall k, \forall m, \forall o$$

The probability conservation constraints are the same as in Eq. 8, while the resource constraints are now associated with (superscripted by) the phase $k$.

Phase-specific binary variables $\Delta^{m,k}_o$ indicate whether or not agent $m$ is assigned one unit of resource $o$ during phase $k$. The constraint $\sum_m \Delta^{m,k}_o = \hat{\omega}_o$ says that the amount of resource $o$ allocated in any phase $k$ must equal the amount of $o$ available (again, we assume a *dummy* agent can hold any unwanted resources). $x^m_{i,a}$ and $\Delta^{m,k}_o$ are linked with the constraint $\frac{\sum_{i \in \mathbb{S}^k} \sum_a u^m_{o,a,i} \times x^m_{i,a}}{T} \leq \Delta^{m,k}_o$ (where $\mathbb{S}^k$ represents the set of states within phase $k$). That is, $x^m_{i,a} \equiv 0$ (i.e., action $a$ is not executable in state $i$ by agent $m$ within phase $k$) if $u^m_{o,a,i} = 1$ (i.e., $a$ requires resource $o$) and $\Delta^{m,k}_o = 0$ (i.e., agent $m$ does not have resource $o$ during phase $k$) for any resource $o$. Otherwise, $x^m_{i,a}$ is not restricted since at most $T$ actions can be executed over the finite time horizon $T$.

Deriving an optimal sequential resource allocation and a joint policy from the solution to Eq. 9 is straightforward. At the start time of phase $k$, resources are redistributed in the following way: if $\Delta^{m,k}_o = 1$, then a unit of resource $o$ is assigned to agent $m$. Every agent $m$ should adopt its policy $\pi^m_{i,a} = \frac{x^m_{i,a}}{\sum_a x^m_{i,a}}$ to maximize the total expected reward of the group of agents.

**Running Example: Optimizing for a Predetermined Phase-Switching Schedule.** *Returning to the running example (Section 4.1), we will see whether the total expected reward can be improved when the resources can be reallocated during execution. Let us say that the predetermined schedule says that resources can be redistributed at times* 1*,* 3*,* 6*, and* 8*, decomposing the example problem's time horizon into four phases of roughly equal duration. Formulating and solving this M-RMP problem with Eq. 9 yields the sequential allocation depicted in Figure 11. Compared to the one-shot distribution (Section 4.2, Figure 10), agent* 2 *no longer idles over the entire horizon, and the total expected reward increases to 65.04,* 31% *higher than attained by the one-shot allocation.*

### 4.6 Determining an Optimal Resource Reallocation Schedule

Without a predetermined resource reallocation schedule, agents can be free (within constraints) to determine for themselves when to reassign resources to achieve their remaining goals better. That is (Section 4.3), given inputs $\mathcal{M}$, $\alpha$, $\mathcal{C}$, and $\mathcal{R}$, the M-RMP optimization problem is to find an optimal resource reallocation schedule (subject to the resource reallocation constraints $\mathcal{R}$), and to find the optimal resource allocation among agents (subject to the resource constraints $\mathcal{C}$) within each phase, as well as to derive optimal executable phase policies for each agent. The complexity of this problem and limitations of straightforward approaches to solving it were described in Section 4.4.

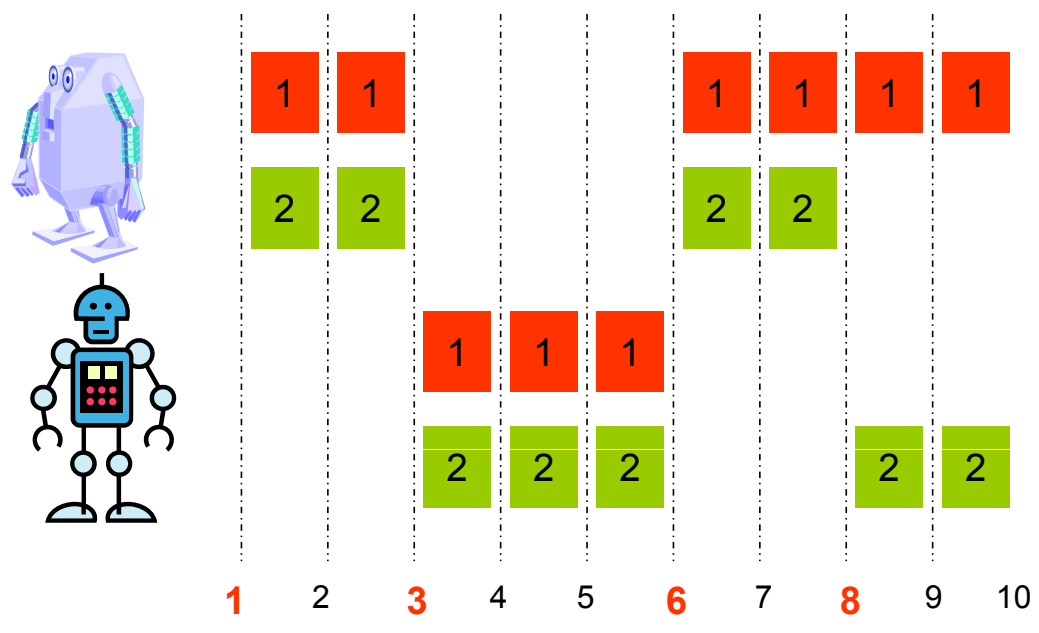

Figure 11: Optimal sequential resource allocation for four predefined phases.

We instead extend the MILP in Eq. 9 to also reason about problem decomposition. The extension is shown in Eq. 10, where (probability conservation) constraints unchanged from Eq. 9 are omitted. To model the constraints on total resource reallocation costs and the occurrences of resource-reallocation events, this new formulation represents the resource constraints at each time step (instead of at each phase), and introduces supplementary constraints to model phase transitions.

$$\max \sum_m \sum_i \sum_a x^m_{i,a} \times r^m_{i,a} \tag{10}$$

subject to:

*probability conservation constraints (unchanged)*

*resource constraints:*

$$\sum_{i \in \mathbb{S}^t} \sum_a u^m_{o,a,i} \times x^m_{i,a} \leq \Delta^{m,t}_o \qquad : \forall t, \forall m, \forall o$$

$$\sum_m \Delta^{m,t}_o = \hat{\omega}_o \qquad : \forall o, \forall t$$

$$\Delta^{m,t}_o \in \{0,1\} \qquad : \forall t, \forall m, \forall o$$

*reallocation constraints:*

$$\Delta^{m,t}_o - \Delta^{m,t-1}_o \leq \Psi_t \qquad : \forall o, \forall t > 1, \forall m$$

$$\Psi_{t=1} = 1$$

$$\sum_t \psi_t \times \Psi_t \leq \hat{\psi}$$

$$\Psi_t \in \{0,1\} \qquad : \forall t$$

where $r^m_{i,a}$, $x^m_{i,a}$, $u^m_{o,a,i}$, $\hat{\omega}_o$, and $T$ have the same definitions as before. $\mathbb{S}_t$ represents the set of states associated with time $t$. New binary variable $\Delta^{m,t}_o$ indicates whether resource $o$ is assigned to agent $m$ at time $t$, and the resource constraints guarantee that the total amount of allocated resources must equal the total amount of available resources at any time point.

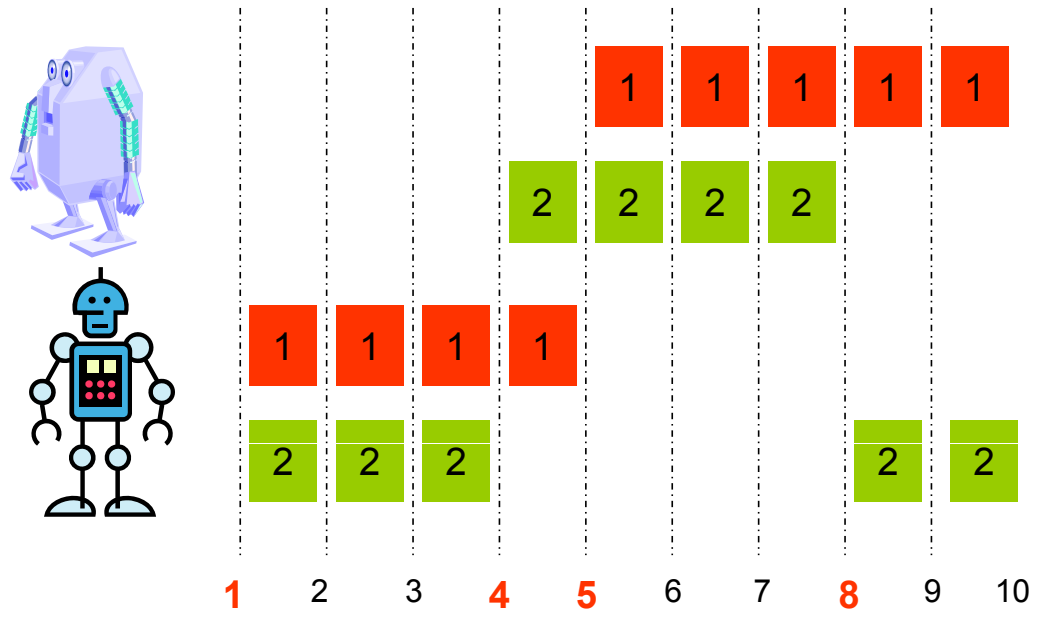

Figure 12: Optimal sequential resource allocation for four phases without a predefined schedule.

To model the cost constraints on resource reallocation, Eq. 10 introduces new binary variables $\Psi_t$ to represent whether the resources are to be redistributed at time $t$. To sidestep the question of what the default resource allocation might be, it is assumed that the resources are always initially allocated at the beginning of the execution, i.e., $\Psi_{t=1} = 1$. Note that $\Delta_o^{m,t} - \Delta_o^{m,t-1}$ can never be greater than one since $\Delta_o^{m,t}$ and $\Delta_o^{m,t-1}$ are binary values in $\{0, 1\}$ ; the constraint $\Delta_o^{m,t} - \Delta_o^{m,t-1} \leq \Psi_t$ thus points out that $\Psi_t$ must be one if any agent $m$ procures any different resource at time $t$ compared to time $t-1$. In other words, any resource reassignment at time $t$ will lead to $\Psi_t = 1$, which means that we can use the constraint $\sum_t \psi_t \times \Psi_t \leq \hat{\psi}$ to limit the total cost for resource reallocation.

By definition, there is a one-to-one mapping between possible sequential resource allocations and possible integer solutions. In addition, given a particular sequential resource allocation, the MILP would be reduced to a linear program whose solution space is equivalent to the executable policy space (because resource constraints would prune unexecutable actions). In other words, the MILP solution space includes the best way of allocating resources together with the best way of utilizing the allocated resources, and so finding an optimal solution to the MILP is equivalent to finding an optimal way of sequentially allocating and utilizing resources.

**Running Example: Optimizing for a Fixed Number of Phase-Switching Times.** *Consider what happens if the agents can determine for themselves a set of reallocation times given an upper bound of four for the size of this set, i.e.,* $\psi_{t=1} = 0$, $\psi_{t\neq 1} = 1$, *and* $\hat{\psi} = 3$. *Using Eq. 10, the optimal schedule to reallocate resources is computed as* $\{1, 4, 5, 8\}$. *Figure 12 depicts the detailed allocation. This schedule gives high priority to and allots sufficient time for agents to accomplish their high-reward tasks (i.e., task* 3 *of agent* 1*, and task* 1 *of agent* 2*). As a result, the total expected reward for the two agents increases to* 72.25*, which is* 11.1% *higher than for the fixed set of 4 reallocation times that were more evenly spaced out (Section 4.5, Figure 11), and* 45.5% *higher than the one-shot case (Section 4.2, Figure 10).*

#### 4.6.1 Variation: Maximizing the Total Reward, Accounting for Cost

As in the S-RMP problem (Section 3.4.1), we can consider a variation of the M-RMP problem where neither the resource-reallocation schedule is predefined (Section 4.5) nor the number of times for reallocating resources is restricted (Section 4.6), but rather that a cost is incurred each time resources are reallocated and this cost is calibrated with the utility of the MDP policy. Thus the optimization problem is to maximize the total expected reward, accounting for the costs of redistributing the resources during execution.

We begin by examining a binary-cost case where, if a resource reallocation is scheduled at time $t$, it will charge the group of agents a constant fee $\psi_t$ regardless of what resources and how many of them are redistributed at that time. In general, coping with such binary reallocation costs is relatively easy because Eq. 10 has paved the way to characterize time steps for resource reassignments.

Eq. 11 shows the changed components of the solution algorithm to this problem, compared to Eq. 10. Eq. 11 adopts a new objective function $\sum_m \sum_i \sum_a x^m_{i,a} \times r^m_{i,a} - \sum_t \psi_t \times \Psi_t$, and removes the constraint $\sum_t \psi_t \times \Psi_t \leq \hat{\psi}$ that is no longer applicable since the agents can now reallocate resources as frequently as they desire.

$$\max \sum_m \sum_i \sum_a x^m_{i,a} \times r^m_{i,a} - \sum_t \psi_t \times \Psi_t \tag{11}$$

subject to:

*probability conservation constraints (unchanged)*

*resource constraints (unchanged)*

*reallocation (cost) constraints:*

$$\Delta^{m,t}_o - \Delta^{m,t-1}_o \leq \Psi_t \qquad : \forall o, \forall t > 1, \forall m$$

$$\Psi_{t=1} = 1$$

$$\Psi_t \in \{0, 1\} \qquad : \forall t$$

Next we consider the more difficult variation where the cost incurred in redistributing resources is based on the *amount* of resources being transferred among the agents. Since it is assumed that the agents are cooperative, it does not matter which agent involved in an exchange pays the resource transfer costs. Without loss of generality, let us say that agent $m$ pays the cost $c^{m,t}_o$ when it obtains one unit of resource $o$ at time $t$ from someone else, and the agent releasing that resource pays no cost.

As before, $\Delta^{m,t}_o$ is used to represent whether resource $o$ is currently held by agent $m$ at time $t$. The cost that agent $m$ should pay for getting resource $o$ at time $t$ can then be represented as $c^{m,t}_o \times \Upsilon(\Delta^{m,t}_o - \Delta^{m,t-1}_o)$ where function $\Upsilon(z)$ is a piecewise linear function, defined as:

$$\Upsilon(z) = \begin{cases} z & z > 0 \\ 0 & \textit{otherwise} \end{cases}$$

This piecewise linear constraint can be equivalently represented using multiple linear constraints by introducing continuous variables $\epsilon^{m,t}_o$. The new MILP formulation is shown in Eq. 12, where again only the groups of changed constraints compared to Eq. 10 are shown.

$$\max \sum_m \sum_i \sum_a x^m_{i,a} \times r^m_{i,a} - \sum_o \sum_m \sum_t c^{m,t}_o \times \epsilon^{m,t}_o \tag{12}$$

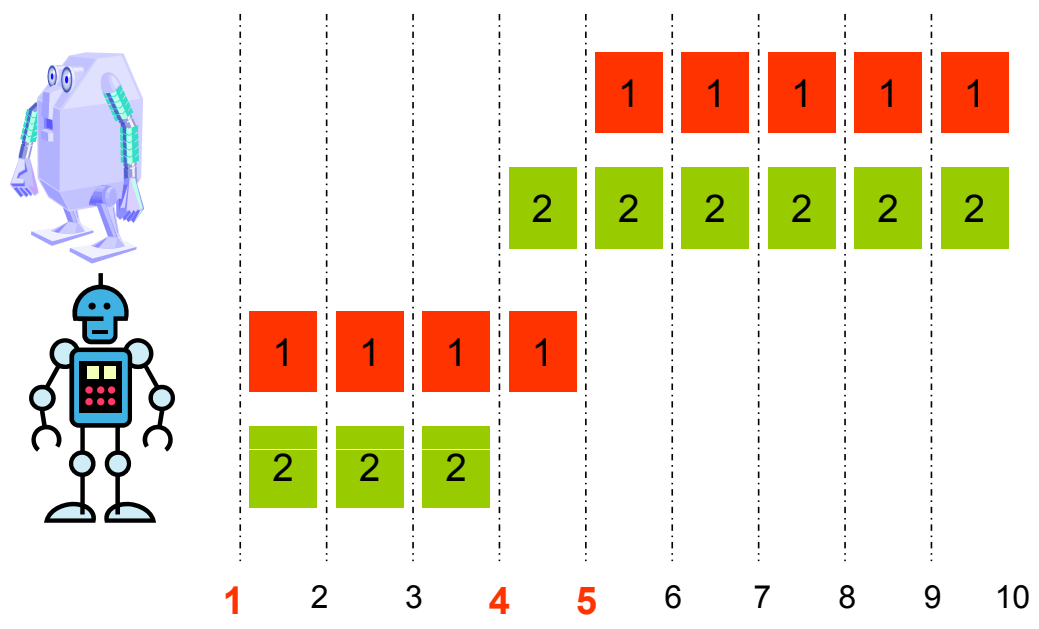

Figure 13: Optimal sequential resource allocation, given that the transfer cost is 5 per unit.

subject to:

*probability conservation constraints (unchanged)*

*resource constraints (unchanged)*

*reallocation (cost) constraints:*

$$\epsilon_o^{m,t=1} = \Delta_o^{m,t=1} \quad : \forall o, \forall m$$

$$\epsilon_o^{m,t} \geq \Delta_o^{m,t} - \Delta_o^{m,t-1} \quad : \forall o, \forall t > 1, \forall m$$

$$\epsilon_o^{m,t} \geq 0 \quad : \forall o, \forall t, \forall m$$

That is, when $t > 1$, $\epsilon_o^{m,t}$ is constrained by $\epsilon_o^{m,t} \geq 0$ and $\epsilon_o^{m,t} \geq \Delta_o^{m,t} - \Delta_o^{m,t-1}$. In other words, $\epsilon_o^{m,t} \geq 1$ when $\Delta_o^{m,t} > \Delta_o^{m,t-1}$ (i.e., $\Delta_o^{m,t} = 1$, and $\Delta_o^{m,t-1} = 0$), and $\epsilon_o^{m,t} \geq 0$ under other circumstances. Note that the objective function of Eq. 12 is to maximize $\sum_m \sum_i \sum_a x_{i,a}^m \times r_{i,a}^m - \sum_o \sum_m \sum_t c_o^{m,t} \times \epsilon_o^{m,t}$, implying the second term $\sum_o \sum_m \sum_t c_o^{m,t} \times \epsilon_o^{m,t}$ should be as small as possible for an optimal solution that yields the highest expected utility. That is, $\epsilon_o^{m,t}$ should reach its lower bound for any optimal solution to Eq. 12, i.e., $\epsilon_o^{m,t} = 1$ when $\Delta_o^{m,t} > \Delta_o^{m,t-1}$ (i.e., when agent $m$ acquires resource $o$ at time $t$) and $\epsilon_o^{m,t} = 0$ otherwise, which exactly matches our expectation of using $\epsilon_o^{m,t}$ to represent the piecewise linear cost function $\Upsilon(\Delta_o^{m,t} - \Delta_o^{m,t-1})$.

**Running Example: Optimizing Total Reward Accounting for Cost.** *Consider how the above algorithm manages the transfer of resources when each transfer incurs cost in the running example problem from Section 4.1. When transferring one unit of any resource costs* 5*, the optimal sequential resource allocation, which is shown in Figure 13, is to transfer only four units of resources over the entire execution (two units at the initial time* 1*, one unit at time* 4*, and one unit at time* 5*). Not surprisingly, as the transfer cost increases, the amount of resources to be transferred decreases, and vice versa.*

*Table 2 compares this resulting schedule with the schedules derived in the previous sections, but now incorporating the cost of* 5 *for each resource transfer. As expected, the algorithm in Eq. 12 yields a reallocation schedule with the highest utility, which is* 48.72*.*

### 4.7 Experimental Evaluation

We analyzed the computational complexity of the M-RMP problem in Section 4.4; here, we empirically evaluate the effectiveness and computational efficiency of the MILP-based solution algorithms we developed in this section, using a grid world environment similar to that used for the S-RMP evaluation in Section 3.5.[14]

| Case | Total Resource (Re)Assignments | Utility |
|---|---|---|
| one-shot (Figure 10) | 2 | 39.64 |
| 4 fixed times (Figure 11) | 7 | 30.04 |
| 3 added times (Figure 12) | 5 | 47.25 |
| unlimited times balancing cost (Figure 13) | 4 | 48.72 |

Table 2: Comparison of the resource reallocation schedules to the example problem, given that the transfer cost is 5 per unit.

#### 4.7.1 Experimental Setup

Each test problem instance includes $m$ cooperative agents where each agent operates in its own $n \times n$ grid world that is independent of all others. The starting location of each agent is always at the center of its grid world.[15] The objective of the group of agents is to maximize their total expected reward within $T$ time steps. Like in the single-agent test problems (Section 3.5), when a grid world is generated, 40% of the locations are randomly chosen as walls, and 10% are randomly chosen as task locations. The rewards of the tasks are randomly set, i.e., the $i^{th}$ task (in a random order) is given reward $i$.

Each task is temporally constrained by its release time and deadline. The release time is the time step when the task becomes available, i.e., attempting the task before then will return zero reward. The deadline is when the task becomes unavailable, i.e., finishing the task after then will also return zero reward. These temporal constraints are randomly set. A task's release time is an integer uniformly and randomly selected in the range $[1,\ T-2]$ where $T$ is the time horizon, and its deadline is always three time steps later. Thus, task $i$'s time window is $[t_i,\ t_i+3)$ where $t_i$ is a random integer in $[1,\ T-2]$. A task can be repeated multiple times (and each time it will give the same reward) within its time window.

The action space of each agent is {*wait*, *up*, *left*, *down*, *right*, *safe-up*, *safe-left*, *safe-down*, *safe-right*, *do*}. All actions except the *do* action have exactly the same definitions as before (Section 3.5). The resource prerequisite of the *do* action is also the same as before, but its outcome no longer always terminates the execution immediately. Instead, it terminates with probability 0.05, and otherwise the agent stays in the same location (with probability 0.95) and can repeat the task or move to another task until the time horizon $T$ is reached. This change makes the test problems more interesting and complex, since each agent now uses resources throughout the experiment rather than just up until completing its first task.

The system is constrained by resource limitations. There are $|O|$ different resource types in the system. Further, there is only one instance of each resource type, which is shared by the $m$ agents.

14. An empirical evaluation in the domain with problems similar to (but more complex than) the running example used in this section (Figure 8) can be found in the work of Wu and Durfee (2007a).
15. Starting in the center makes the problem more interesting and challenging than starting in a corner, allowing the agent to potentially visit a larger fraction of the grid world sooner.

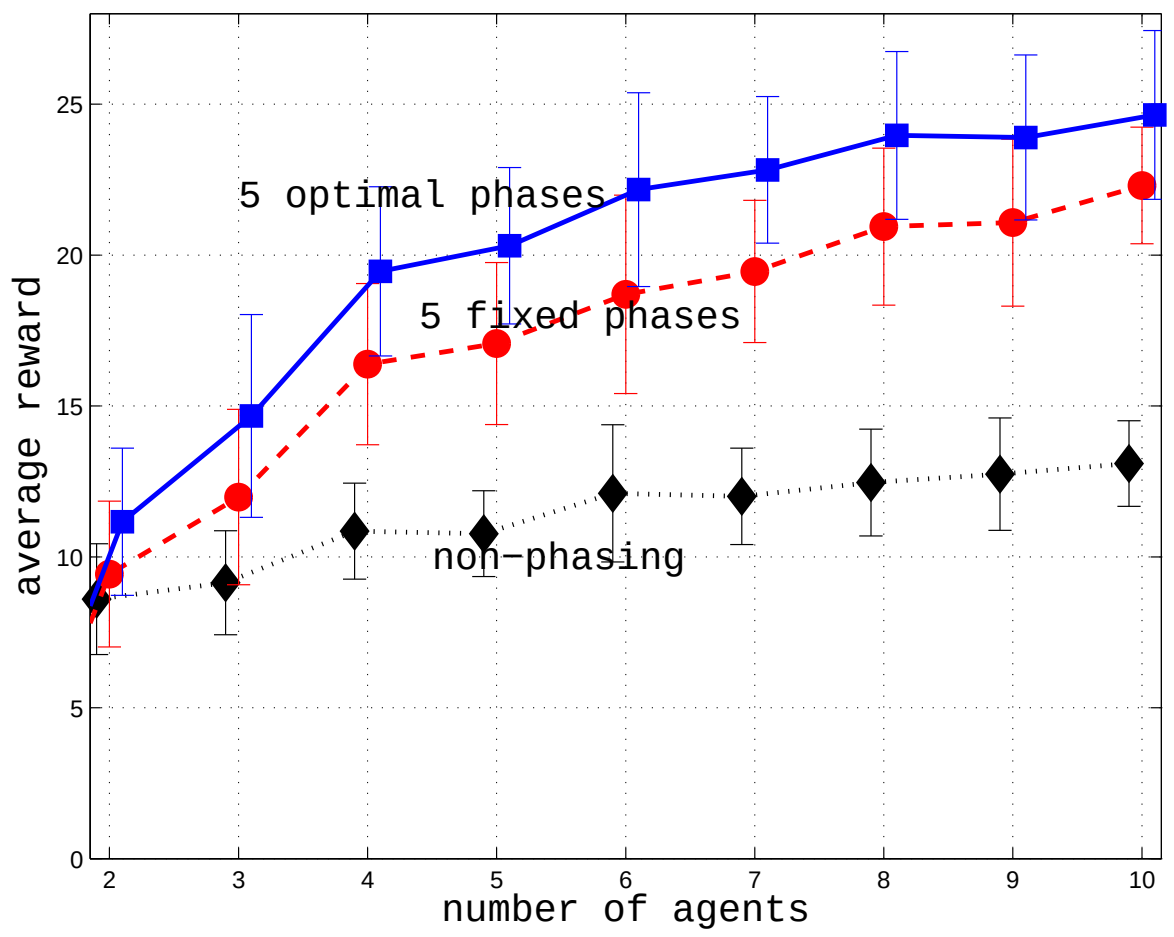

Figure 14: Exploiting fixed phases increases the reward, and finding optimal phases further increases the reward.

#### 4.7.2 Improvements to Solution Quality

Figure 14 demonstrates the improvement of our sequential resource allocation approaches over the prior one-shot resource allocation approach. The $x$-axis of the figure represents the number of agents in the world, and the $y$-axis specifies the total expected reward of the group of agents.[16] Other parameters are set as follows: $T = 10$, $n = 5$, and $|O| = 5$. We can see that, by taking into account resource reallocation opportunities during execution, the agents can gain a considerably higher reward. For example, in the case that five fixed resource (re)allocation times (one at the initial time step and the other four randomly and uniformly selected when the test problem is defined) are available in the midst of execution, our mission-phasing approach, using Eq. 9 and denoted as *5-fixed-phases*, on average achieves a reward 50% higher than that of not exploiting resource-reallocation opportunities. We can also see that (as expected) finding and using the optimal resource-allocation and phase-switching time points can further improve the system performance, e.g., the *5-optimal-phases* approach (using Eq. 10 and assuming that four additional phase-switching points besides the one at the initial time step can be created) achieves an average reward about 20% higher than the aforementioned 5-fixed-phases solution.

Another interesting observation from Figure 14 is that the improvement of sequential resource allocation over one-shot resource allocation increases as the number of agents increases. This is because, given that the number of resources is fixed at 5, the more agents there are, the scarcer the resources are. Hence, assigning a resource to the right agent at the right time becomes increasingly important to performance as the constrainedness of the system increases.

Figure 15 uses the same parameters as Figure 14 (i.e., $T = 10$, $n = 5$, and $|O| = 5$), but now holds the number of agents constant at $m = 5$, and shows how much better agents that can choose phase-switching times (Eq. 10) can do as the number of phase-switching times

16. In this section, each average data point is computed from 20 random test problems.

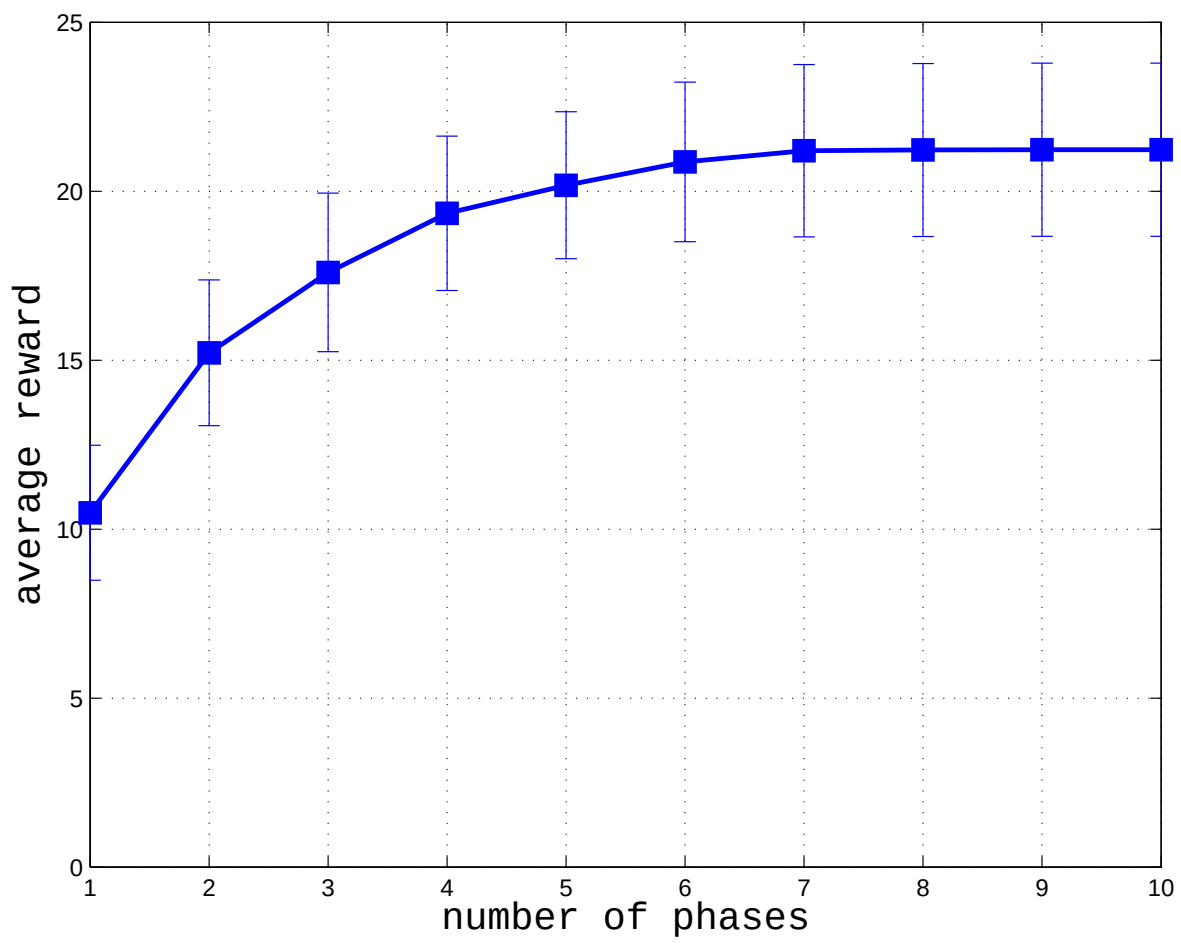

Figure 15: The reward increases as the phase-switching cost limit (that determines the number of phases) increases.

allowed rises. However, note that, unlike the S-RMP optimization problem, even if the agents can reallocate resources at every time step, they usually cannot achieve the same reward as in the unconstrained case with unlimited resources (which has a reward of 37.2 on average in these test problems). This is because, at each time step, some agents might not be able to acquire their most desired resources because there simply are not enough of these resources to go around.

### 4.7.3 Computational Efficiency

To understand the impact of the number of phases on the computational cost and to choose "hard" M-RMP test problems for the computational efficiency evaluation to follow, we now run experiments with the same parameters as in Figure 15, but collect and examine the results of average runtime for finding exact solutions to the test problems. As shown in Figure 16, the MILP-based solution approach can exploit over-constrainedness (when the number of phases is small) and under-constrainedness (when the number of phases is large) to improve efficiency. This complexity profile indicates that on average, with these parameter settings, the problems are most difficult when constrained to 3 phases (that is $\hat{\psi} = 2$ and as usual $\psi_{t=1} = 0$). In the experiments that follow, one parameter is varied at a time while the others retain their default settings, and these variations create larger and more complex instances than used to generate Figure 16. Hence, the phase-switching cost limit $\hat{\psi}$ is by default set to 3 to avoid simpler over-constrained problems.

We compare our MILP-based algorithm (Eq. 10) with the WDP-based algorithm (using the auction-based resource allocation strategy), which is the most computationally-efficient approach among the three prior related approaches discussed in Section 4.4. Recall that the WDP-based algorithm involves two steps. First, each agent submits valuations of its possible sequential resource allotments to a central agent. The number of bids is $C_{K-1}^{T-1} \times (2)^{|O| \times K}$ (as explained in Section 4.4). Second, the central agent solves a winner determination problem.

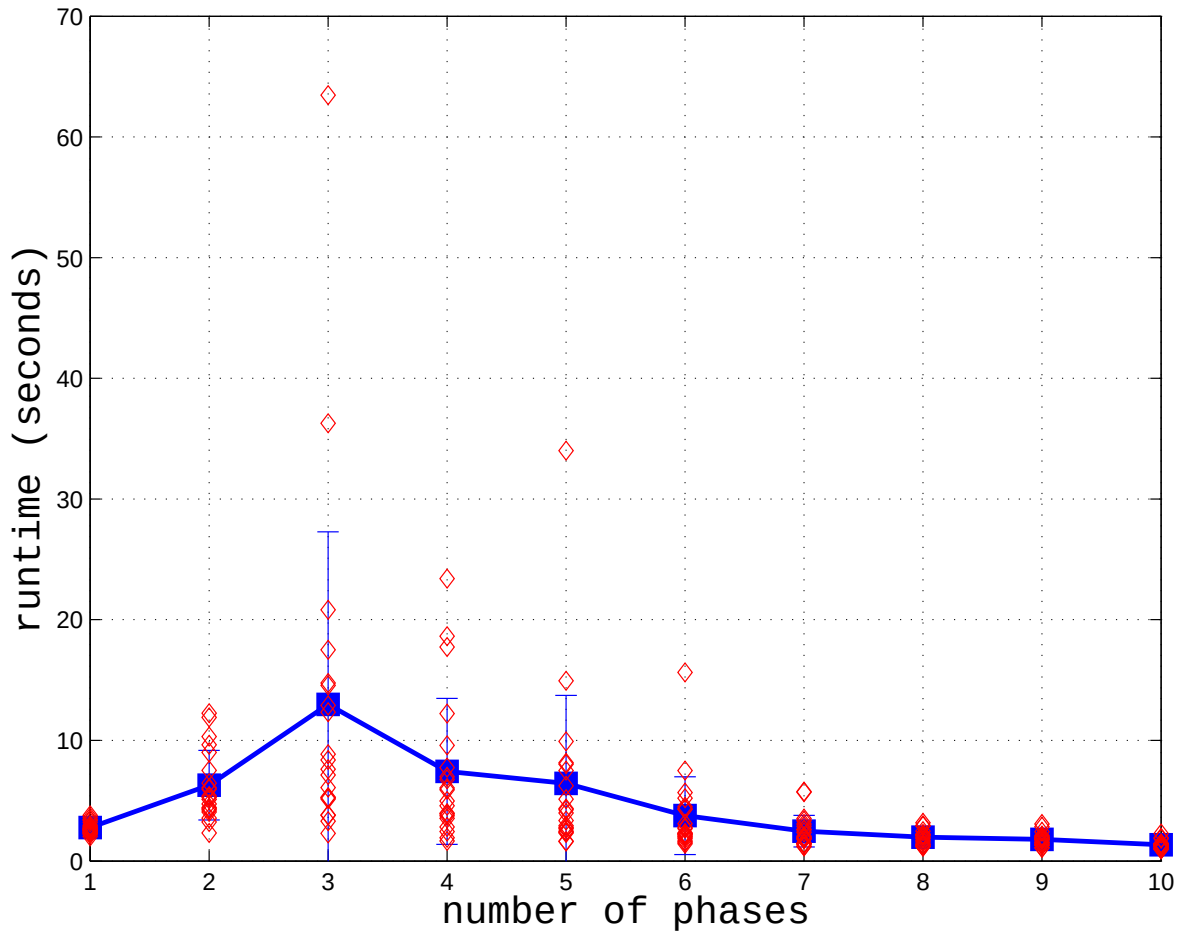

Figure 16: The runtime increases and then decreases as the number of times of resource reallocation increases.

Let us assume that the central agent has a perfect filtering method (although it usually does not), and so it only needs to consider and evaluate "valid" combinations of bids. This assumption reduces the number of possible combinations from $(C_{K-1}^{T-1} \times (2)^{|O| \times K})^m$ to $C_{K-1}^{T-1} \times (m)^{|O| \times K}$ where the base $m$ in the exponentiation $(m)^{|O| \times K}$ is because there are $m$ different ways to allocate one resource among $m$ agents.

However, even with this enhancement, the WDP-based algorithm is still computationally intractable for moderately complex M-RMP problems (where it often cannot find an exact solution within 100 hours of cpu time). Note that the lower bound of the running time of the WDP-based algorithm can be approximated as $C_{K-1}^{T-1} \times (2)^{|O| \times K} \times t_{bid} + C_{K-1}^{T-1} \times (m)^{|O| \times K} \times t_{eval}$ where $t_{bid}$ is the average runtime of evaluating a sequential resource allotment (i.e., a bid) by modeling and solving an unconstrained finite-horizon MDP, and $t_{eval}$ is the average runtime of evaluating a feasible combination of agents' bids. This work uses a sampling method to estimate the runtime, i.e., $t_{bid}$ and $t_{eval}$ are estimated from 100,000 random runs.

Figure 17 compares the average runtime results under various parameter settings.[17] Note that the $y$-axis is in a logarithmic scale. These results illustrate and emphasize that the MILP-based algorithm, which formulates and simultaneously solves the coupled problems of mission decomposition, resource allocation, and policy formulation using a single compact MILP formulation, can effectively and fruitfully exploit the inter-relationships among these component problems. As a result, it is significantly faster than the WDP-based approach that considers the component problems in isolation.

## 4.8 Summary

In this section, we have presented, analyzed, and empirically evaluated an MILP-based approach that automates the process of finding and using optimal resource reallocation schedules

17. Neither MILP nor WDP uses parallel computation.

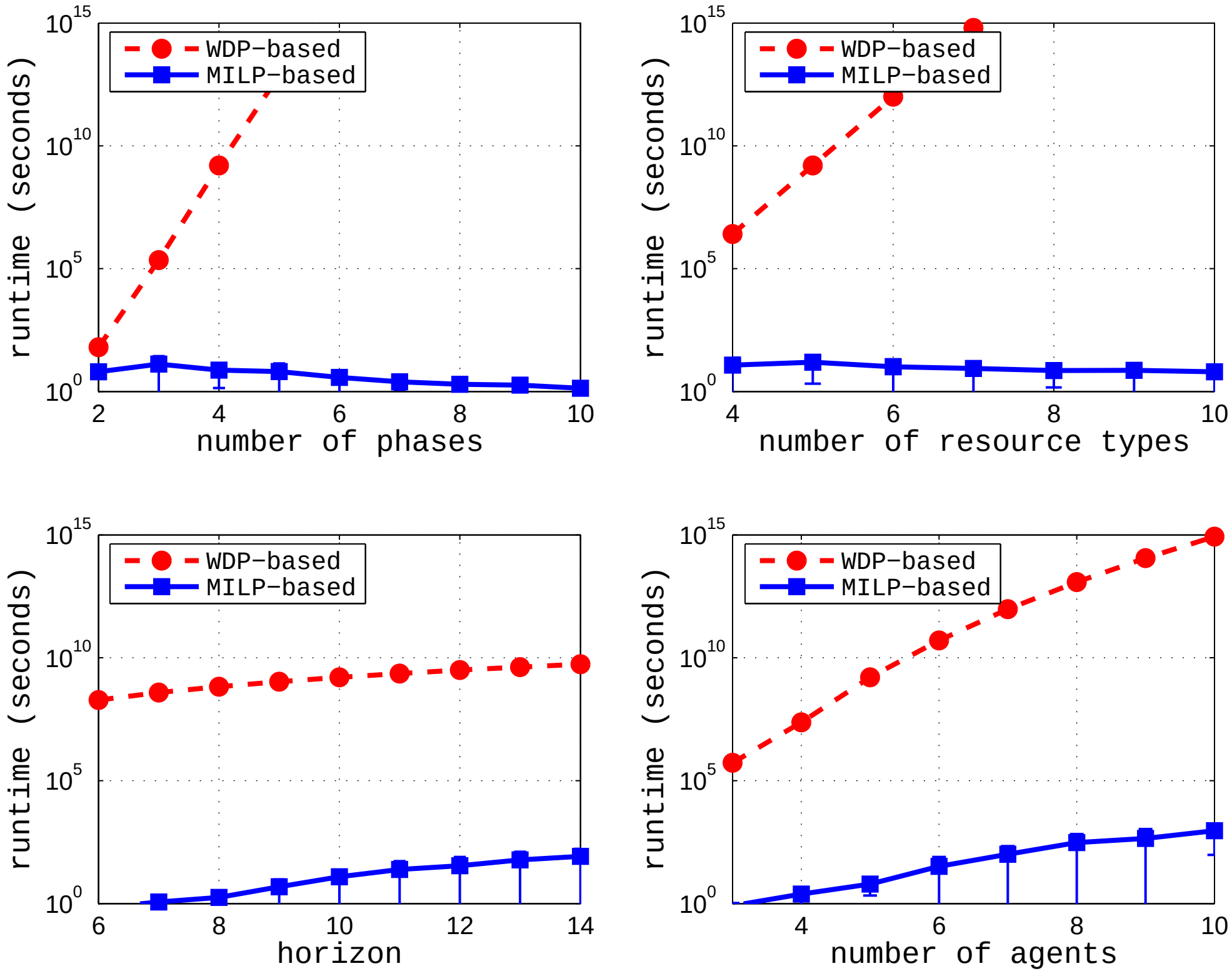

Figure 17: Runtime comparison between the MILP-based algorithm and the WDP-based algorithm. Parameters are set as follows: Top-left figure $n = 5$, $T = 10$, $m = 5$, $|O| = 5$, and $\hat{\psi} = \{1, 2, ..., 9\}$. Top-right figure $n = 5$, $T = 10$, $m = 5$, $|O| = \{4, 5, ..., 10\}$, and $\hat{\psi} = 3$. Bottom-left figure $n = 5$, $T = \{6, 7, ..., 14\}$, $m = 5$, $|O| = 5$, and $\hat{\psi} = 3$. Bottom-right figure $n = 5$, $T = 10$, $m = \{3, 4, ..., 10\}$, $|O| = 5$, and $\hat{\psi} = 3$.

for a group of agents operating in complex environments with resource limitations and with uncertainties. Our analytical and experimental results have shown that the approach can greatly reduce computational cost compared to prior approaches.

## 5. Related Work

The resource-driven mission phasing (RMP) problem involves three intertwined component problems: mission (problem) decomposition, resource configuration, and policy formulation. Each of these component problems has been studied in a wide variety of research fields. The combinations of any two of them have also gained much attention in recent years. This section gives an overview of related work, and discusses why those prior approaches are not directly applicable to the RMP problem of interest in this paper.

As was presented in Section 3.1 and Section 4.3, the RMP problems are defined by extending an unconstrained MDP model to include resource constraints and phase-switching constraints. The organization of this section follows the way of that definition. It begins with a discussion of policy formulation techniques, followed by a discussion of resource configuration techniques. It then reviews problem decomposition techniques and their combinations with policy formulation and/or resource configuration work. This section concludes with a discussion of the "mode-transition" research that is related to this work but does not fit clearly into the previous categories.

### 5.1 Policy Formulation.

The well-known Markov decision process has been described in Section 2.1. By formulating a sequential decision-making problem into a MDP model, a number of efficient (polynomial-time) solvers, such as the value iteration and policy iteration algorithms, can be used to compute an optimal policy (Puterman, 1994).

However, directly applying these algorithms in resource-constrained systems, such as the resource-driven mission-phasing problem, typically involves incorporating resource features in the MDP state representation (and so actions can be conditioned on resource availability), which will result in an exponential increase in the size of the state space (Meuleau et al., 1998), i.e., the well known "curse of dimensionality" challenge. It has been shown in the empirical results (Section 3.5.3) that the exponential-size state space can result in computational inefficiency.

### 5.2 Resource Configuration.

Because in some domains it is impossible (or expensive) to resolve resource constraints by modifying the agent's physical architecture (for example, adding another battery to a robot already deployed on Mars), improving the performance of a constrained agent under its limited architecture has been an active subject in recent years, i.e., a class of "bounded optimality" problems (Russell, 2002). The Cooperative Intelligent Real-Time Control Architecture (CIRCA) is one such research effort (Musliner, Durfee, & Shin, 1993, 1995). CIRCA uses a simple greedy, myopic approach to compute feasible policies. It starts with building an optimal unconstrained policy without worrying about its real-time requirements, and then greedily repairs the policy until executable on the real-time system.

Not surprisingly, the (fast) greedy approach adopted by CIRCA might result in suboptimal policies that cannot fully utilize the agent's capacity. Several other recent studies have proposed alternative algorithms for searching for a policy that is executable within the agent capacity constraints and that optimizes the expected (possibly discounted) reward accrued over the entire agent execution. For example, Altman (1998) adopted a Lagrangian and dual LP approach to solve constrained MDPs with total cost criteria. Feinberg (2000) analyzed the complexity of constrained discounted MDPs. Of particular relevance to the work in this paper is the study of strongly-coupled resource allocation and policy formulation problems by Dolgov and Durfee (2006). Their approach implements simultaneous combinatorial optimization and stochastic optimization via reduction to mixed integer linear programming, which has been recapped in Section 2.2. However, these prior studies on constrained agents are based upon the assumption that the agent's limited capacity is configured by the resources it procures prior to execution but cannot be reconfigured during plan execution.

### 5.3 Problem Decomposition.

In the literature of stochastic planning, a number of decomposition algorithms have been proposed to speed up the planning process. The discovery of "recurrent classes" of MDPs is one such decomposition strategy, which can find an exact state space decomposition in an environment with uncertainties (Puterman, 1994; Boutilier, Dean, & Hanks, 1999). A recurrent class represents a special *absorbing* subset of the state space, which means that once an agent enters a recurrent class it remains there forever no matter what policy it adopts. Puterman (1994) has suggested a variation of the Fox-Landi algorithm (Fox & Landi, 1968) to discover recurrent classes. With the discovery of the recurrent classes, the MDP solver can derive an optimal overall policy by building an optimal policy in each recurrent class independently and then constructing and solving a reduced MDP consisting only of transient states (i.e., removing the recurrent classes in the MDP).

Of course, not all application problems can be exactly decomposed into independent subproblems. However, many of them are composed of multiple weakly-coupled sub-problems where the number of states and transitions connecting two neighboring sub-problems is relatively small. A number of heuristic decomposition methods have been designed to exploit such weakly-coupled relationships. As an example, in the robot navigation domain (Parr, 1998; Precup & Sutton, 1998; Lane & Kaelbling, 2001), doorways (or similar connection structures, such as bridges) can be used to break a large environment into blocks of states, e.g., one block for each room. Two neighboring blocks are only connected by a small number of *doorway* states. Once a weakly-coupled state space is decomposed into several pieces, there are a few methods that can be used to efficiently build an overall policy based upon sub-problem policies. One common method is to let each sub-problem iteratively exchange information with its neighboring sub-problems, and repeatedly revise its sub-policy (if necessary) based upon its updated knowledge about utilities or values of its neighbors until an overall (approximately) optimal solution is derived (Dean & Lin, 1995).

Besides the application in stochastic planning, decomposition techniques have also been shown to be beneficial for resource management in many realistic application domains. Several resource allocation algorithms have been developed for the problem of allocating a set of heterogeneous resources with availability constraints to maximize a given utility function (Wu & Castanon, 2004; Palomar & Chiang, 2006; Reveliotis, 2005). For example, Wu and Cas-tanon (2004) presented an approximate solution algorithm using decomposition combined with dynamic programming, and their experimental results showed that the algorithm produces near-optimal results with much reduced computational effort.

In addition to the Artificial Intelligence (AI) techniques discussed above, decomposition techniques, which are often integrated with hierarchical control (also called multilevel control in some literature), have received much attention in recent years in Operations Research, Operations Management, Systems Theory, Control Theory, and several other fields (Sethi, Yan, Zhang, & Zhang, 2002; Antoulas, Sorensen, & Gugercin, 2001; Xiao, Johansson, & Boyd, 2004; Phillips, 2002; Teneketzis, Javid, & Sridhar, 1980). Many manufacturing systems are large and complex; the management of such systems requires recognizing and reacting to a wide variety of events that could be deterministic or stochastic. Obtaining exact optimal policies to run these systems is often very difficult both theoretically and computationally. By exploiting the fact that real-world systems are often characterized by several decision subsystems, e.g., a company consists of departments of marketing, production, personnel, and so on, one popular way to deal with the computational complexity challenge is to develop methods of hierarchical decision-making for these systems. The fundamental ideas are to reduce the overall complex problem into multiple smaller, manageable sub-problems, to solve these sub-problems, and to coordinate solutions to the sub-problems so that overall system objectives and constraints are satisfied (Sethi et al., 2002).

To summarize, it is well established that utilizing decomposition can greatly reduce computational costs in many situations. However, all the aforementioned prior decomposition techniques are not directly applicable to the RMP optimization problem. The underlying reason is that decomposition points that are good at reducing computational efforts are not necessarily (and possibly completely unrelated to) the optimal points for constrained agents to reconfigure resources. It is worth emphasizing that RMP decomposition tackles capacity constraints instead of computation time constraints. Indeed, in general, the mission decomposition in the RMP solution will not in itself reduce computational requirements because a policy in one phase can usually only be optimized with respect to the policies planned for possible subsequent phases.

### 5.4 Mode Transition.

Finally, it is important to distinguish the resource-driven mission-phasing research from the "mode-transition" research implemented in the fields of Operations Research and Control Theory (Schrage & Vachtsevanos, 1999; Wills, Kannan, Sander, Guler, Heck, Prasad, Schrage, & Vachtsevanos, 2001; Karuppiah, Grupen, Hanson, & Riseman, 2005). At first glance, these two research fields have a lot in common: they both work on transitions from one sub-problem to another, and both take into account resource reconfigurations. However, it should be pointed out that they emphasize distinct aspects, and are applicable to different application domains.

First of all, in the mode-transition approach, operational modes are usually tightly associated with some explicit actions (e.g., *hover* and *fly-forward* modes in the helicopter example described by Schrage & Vachtsevanos, 1999), corresponding to some particular states (e.g., *sleep*, *search*, *seed*, and *final* modes defined by Bojinov, Casal, & Hogg, 2002), or characterized with some explicit purposes (e.g., passing through a narrow tunnel and then traversing rough terrain requires a self-reconfiguring robot to adjust its shape to achieve its goal better, Rus

& Vona, 2001). In contrast to the explicit definition or representation of modes in the mode-transition research, phases in the RMP problem are usually much more difficult to identify. The phasing information is hidden in the MDP model, and finding optimal phases is usually a challenging task.

Second, in the mode-transition research, mode transition and resource reconfiguration are often triggered by real-time events, e.g., responding to an unexpected disastrous event and reconfiguring resources for fault toleration (Drozeski, 2005). In contrast, the resource-driven mission-phasing study assumes that a decision-making agent has complete information about the environment prior to its execution, and one of its main objectives is to find the optimal points for reconfiguring resources and capacity usage. That is, phase switching in RMP is a choice of the agent instead of a reactive response to an exogenous event. More specifically, the RMP techniques presented in this paper utilize sequential decision-making to identify optimal resource reconfiguration and policy switching states. They emphasize how to reconfigure resources and switch policies so that the agent(s) would not (or would be less likely to) enter into the predicament of encountering undesirable events, instead of studying how to reconfigure resources in real-time to respond to an unexpected event.

Finally, much prior mode-transition research, particularly in the Control Theory literature, investigates how to perform a smooth functional transition among modes, but the work in this paper simply assumes that there are *aggregate* resource (re)configuration actions, each of which can be a sequence of primitive actions of arranging resources. This paper does not address the details of how the agents *mechanically* implement mode-transition and resource-reconfiguration actions.

## 6. Conclusion

The work in this paper designed, analyzed, and evaluated a suite of computationally efficient algorithms that can automatically identify and utilize resource reconfiguration opportunities in resource-constrained environments. The analytical and experimental results illustrated and emphasized that the mission phasing approach, incorporating problem decomposition, resource allocation, and policy formulation, can help constrained agents judiciously and effectively exploit resource reconfiguration opportunities to improve their performance.

This section concludes the paper with a summary of the main contributions of this work and a discussion of several promising future research directions.

### 6.1 Summary of Contributions

- ▷ This work explicitly took into account known opportunities in the midst of execution to reconfigure resources and switch policies, and designed computationally efficient algorithms (including an abstract MDP algorithm for single-agent resource reconfiguration problems and a MILP-based algorithm for multiagent resource reallocation problems) to optimize the use of these fixed opportunities in complex stochastic systems. The empirical results (Figure 4 and Figure 14) confirmed that exploiting such phase-switching opportunities can considerably improve performance, particularly in tightly constrained systems (the reward doubles in some test cases).

- ▷ As an extension to utilizing fixed phase-switching opportunities, Section 3.4 (for single-agent systems) and Section 4.6 (for multiagent systems) presented MILP-based algo-rithms that are able to automate the process of finding and using mission phases in stochastic, constrained systems, which not only eliminates the need for having phases predefined, but also avoids potential sub-optimality caused by phases being improperly predefined.

- ▷ The automated resource-driven mission-phasing algorithms presented in this work are computationally efficient. By capturing a whole mission-phasing problem into a compact mathematical formulation and then simultaneously solving the coupled problems of mission decomposition, resource allocation, and policy formulation, the presented algorithms can effectively exploit problem structure, which results in a significant reduction in computational cost in comparison with an approach that considers mission decomposition, resource allocation, and policy formulation in isolation (e.g., a reduction from hours to seconds as was shown in Figure 17).

- ▷ Unlike much prior work where agents reactively (and often greedily) reconfigure resources when exogenous events occur, this work, based upon Markov decision processes and sequential decision-making theory, can proactively determine and optimally utilize resource reconfiguration opportunities. It provides a new computationally efficient resource-reconfiguration mechanism for resource-constrained environments.

### 6.2 Future Work

Although this paper presented a suite of algorithms to improve agent performance in constrained stochastic systems, there is still much interesting work remaining. Below, we point out a few promising research directions to overcome some of the limitations of the work we have presented in this paper.

- ▷ **Resource Constraints and Time Limitations**

  Resource-driven mission-phasing problems are NP-complete. Although the solution approaches designed in this work can exploit problem structure to reduce computational cost, finding an exact solution to a complex RMP problem might still be difficult, particularly in time-limited environments. One approach to handling such problems is to adopt approximate methods. Our preliminary investigations into developing anytime algorithms for solving problems with both resource constraints and time limitations has shown promise (Wu, 2008) but more work remains in this area, including comparing methods grounded in the RMP concepts with heuristic and greedy techniques for allocating resources to agents.

- ▷ **More Flexible Resource Reallocation Options**

  The work we discussed in this paper assumed a clear delineation between two kinds of states: states where resources can be (re)allocated in any way desired (phase-switching states) and states where resource allocations cannot change. More generally, it might be the case that states could exist where only limited resource reallocations could occur (e.g., a partially filled "toolbox", or a subset of other agents with whom to swap resources), leading to more challenging reasoning problems for agents to decide which such states to seek out and avail themselves of. Further, an agent could even potentially create such a state on the fly by dropping off resources in a well-chosen state to be gainfully retrieved (perhaps by another agent) at a future time. Obviously, as problems get increasingly complicated in these kinds of ways, modeling resources as part of state and incorporating actions of picking up and dropping off resources becomes important, leading to MDP-based solution techniques as described in this paper (e.g., Sections 3.2 and 3.5.3). Finding methods for having more of this kind of flexibility without incurring the costs of the MDP-based techniques is a challenging direction for future work.

▷ **Resource Reallocation and Decentralized MDPs**

A limiting assumption made in this work is that, once a resource reallocation is scheduled, participant agents will always be able to successfully redistribute resources among themselves at that scheduled time, regardless of what their other state features' values are. We plan to relax this assumption in the future to consider sequential resource allocation problems with additional constraints on when and where the agents are able to exchange resources. For example, physical agents might only be able to exchange resources when they are at the same location at the same time. Or, as another example, a task might not be interruptible once it has started, which means that it may be impossible to reassign the resources used by that task until the task has completed.

Decentralized MDPs are one possible way to solve such problems. Some of our preliminary work (Wu & Durfee, 2006), which is not included in this paper, has developed a MILP-based algorithm for solving transition independent Dec-MDPs. That work linked the Dec-MDP formulation with the MILP formulation, and pointed out one way to characterize resource constraints in the MILP formulation. In the future, we will dig deeper in this direction.

▷ **Application and Evaluation in Other Settings**

Our work has so far focused on testing our techniques on problems that are small enough to solve using slower standard approaches (to confirm that our techniques are optimal) and on generated spaces of problems that allow us to probe the efficacy of our techniques in settings that vary in controlled ways. Applying our techniques to more extensive and realistic domains is an important avenue to follow to identify their strengths and weaknesses better. An example application that we are particularly interested in, and that prompted this work in its early stages, is intelligent real-time control. Systems such as CIRCA (Musliner et al., 1993, 1995) use AI techniques to construct real-time control plans that are composed of a set of sense-act tasks that are scheduled at frequencies to ensure safe operation. Often, all of the desired sense-act tasks cannot fit into the schedule, so the most important combination of tasks must be chosen. In applications such as controlling an unmanned aircraft (Atkins, Abdelzaher, Shin, & Durfee, 2001), different combinations might be better in different phases of activity (takeoff, cruising, landing, etc.). Similarly, for a formation of aircraft, which aircraft is responsible for detecting and reacting to a particular event can shift as time progresses in the mission (Musliner, Goldman, & Krebsbach, 2005). While prior work has used heuristic, local search techniques to find good solutions to such phasing problems, the techniques in this paper have the potential of finding optimal mission decompositions.

## Acknowledgments

This material is based upon work supported in part by the DARPA/IPTO COORDINATORs program and the Air Force Research Laboratory under Contract No. FA8750–05–C–0030, and by the Air Force Office of Scientific Research under Contract No. FA9550-07-1-0262. The views and conclusions contained in this document are those of the authors, and should not be interpreted as representing the official policies, either expressed or implied, of the Defense Advanced Research Projects Agency, the Air Force, or the U.S. Government.

The authors thank Dmitri Dolgov and the three anonymous reviewers for their very helpful suggestions and comments, and Stefan Witwicki and Jim Boerkoel for their help in proofreading this article.